\documentclass[fleqn,usenatbib]{mnras}

\usepackage{newtxtext,newtxmath}

\usepackage[T1]{fontenc}

\DeclareRobustCommand{\VAN}[3]{#2}
\let\VANthebibliography\thebibliography
\def\thebibliography{\DeclareRobustCommand{\VAN}[3]{##3}\VANthebibliography}

\usepackage{graphicx}	
\usepackage{amsmath}	

\newcommand{\hMpc}{h^{-1}{\rm Mpc}}
\newcommand{\hMsun}{h^{-1}{\rm M}_\odot}

\newcommand{\teal}[1]{\noindent \textcolor{black}{{{#1}}}}

\title[Mass and environment effects on SZ with MTNG]{Interpreting Sunyaev-Zel'dovich observations with MillenniumTNG: Mass and environment scaling relations}

\author[B. Hadzhiyska et al.]{%
Boryana Hadzhiyska,$^{1,2,3}$\thanks{E-mail: boryanah@berkeley.edu}
Simone Ferraro,$^{3,2}$
R\"udiger Pakmor,$^{4}$
Sownak Bose,$^{5}$
Ana Maria Delgado,$^{6}$
\newauthor%
C\'esar Hern\'andez-Aguayo,$^{4,8}$
Rahul Kannan,$^{7}$
Volker Springel,$^{4}$
Simon D. M. White,$^{4}$
and Lars Hernquist$^{6}$
\vspace*{0.1cm}\\%
$^{1}$Miller Institute for Basic Research in Science, University of California, Berkeley, CA 94720, USA\\%
$^{2}$Berkeley Center for Cosmological Physics, Department of Physics,
University of California, Berkeley, CA 94720, USA\\%
$^{3}$Lawrence Berkeley National Laboratory, One Cyclotron Road, Berkeley CA 94720, USA\\%
$^{4}$Max-Planck-Institut f\"ur Astrophysik, Karl-Schwarzschild-Str. 1, D-85748, Garching, Germany\\%
$^{5}$Institute for Computational Cosmology, Department of Physics, Durham University, South Road, Durham, DH1 3LE, UK\\%
$^{6}$Center for Astrophysics | Harvard $\&$ Smithsonian, 60 Garden St, Cambridge, MA 02138, USA\\%
${^7}$Department of Physics and Astronomy, York University, 4700 Keele Street, Toronto, ON M3J 1P3, Canada\\%
$^{8}$Excellence Cluster ORIGINS, Boltzmannstrasse 2, D-85748 Garching, Germany
}

\date{Accepted XXX. Received YYY; in original form ZZZ}

\pubyear{2015}

\begin{document}
\label{firstpage}
\pagerange{\pageref{firstpage}--\pageref{lastpage}}
\maketitle

\begin{abstract}
In the coming years, Sunyaev-Zel'dovich (SZ) measurements can dramatically improve our understanding of the Intergalactic Medium (IGM) and the role of feedback processes on galaxy formation, allowing us to calibrate important astrophysical systematics in cosmological constraints from weak lensing galaxy clustering surveys. However, the signal is only measured in a two-dimensional projection, and its correct interpretation relies on understanding the connection between observable quantities and the underlying intrinsic properties of the gas, in addition to the relation between the gas and the underlying matter distribution. One way to address these challenges is through the use of hydrodynamical simulations such as the high-resolution, large-volume MillenniumTNG suite. \teal{We find that measurements of the optical depth, $\tau$, and the Compton-y parameter, $Y$, receive large line-of-sight contributions which can be removed effectively by applying a Compensated Aperture Photometry (CAP) filter. In contrast with other $\tau$ probes (e.g., X-rays and Fast Radio Bursts), the kSZ-inferred $\tau$ receives most of its signal from a confined cylindrical region around the halo due to the velocity decorrelation along the line-of-sight. Additionally, we perform fits to the $Y-M$ and $\tau-M$ scaling relations and report best-fit parameters adopting the smoothly broken power law (SBPL) formalism. We note that subgrid physics modeling can broaden the error bar on these by 30\% for intermediate-mass halos ($\sim$$10^{13} \, {\rm M}_{\odot}$). The scatter of the scaling relations can be captured by an intrinsic dependence on concentration, and an extrinsic dependence on tidal shear. Finally, we comment on the effect of using galaxies rather than halos in real observations, which can bias the inferred SZ profiles by $\sim$20\% for $L_\ast$-galaxies.}

\end{abstract}




\section{Introduction}
\label{sec:intro}

In the next few decades, cosmological observations across all wavelengths will see a significant increase in their size and scope. In particular, galaxy surveys such as the Dark Energy Spectroscopic Instrument  \citep[DESI,][]{2016arXiv161100036D}, the Vera Rubin Observatory \citep[\textit{Rubin,}][]{0912.0201} and the Nancy Grace Roman Space Telescope \citep[\textit{Roman,}][]{2015arXiv150303757S} will map out the late-time structure of the Universe in three dimensions and address a number of long-standing questions, among which are the equation of state of dark energy and the growth rate of structure across cosmic time. Meanwhile, high-resolution measurements of early-Universe probes such as the cosmic microwave background (CMB) will shed light on various aspects of both cosmology and astrophysics. 

The analysis of primary CMB anisotropies dating back to the surface of last scattering, yields invaluable insights into the energy content of the Universe, its curvature and age, and the nature of the primordial processes that seeded the CMB fluctuations. On the other hand, the secondary CMB anisotropies, imprinted during the CMB photon's long journey towards the observer, contain detailed information about the distribution of (baryonic and non-baryonic) matter along the line of sight. Studying the statistics of that distribution can reveal the answers to fundamental physics puzzles such as the sum of the neutrino masses as well as the nature of dark energy, gravity, and the primordial Universe \citep{2018PhRvD..98h3501V,2018JCAP...09..017C}. Prominent examples of late-time processes affecting CMB photons are the thermal and kinematic Sunyaev-Zel'dovich (tSZ and kSZ) effects, which are also the focus of this study.

The tSZ effect is caused by the inverse Compton scattering of CMB photons with the hot electrons in galaxy clusters and, as such, has many potential uses in cosmology and astrophysics. Through tSZ, which is sensitive to a wide range of redshifts, we can detect massive clusters at various epochs \citep[e.g.,][]{2013ApJ...763..127R}. The integrated tSZ signal of individual clusters, known as the Compton-y parameter, can serve as a mass proxy via the use of scaling relations \citep{2015ApJ...806...18C,2019A&A...626A..27S}, and redshift-calibrated cluster counts can constrain the shape of the halo mass function on the high-mass end, which is highly sensitive to the amplitude of density fluctuations, the sum of the neutrino masses, and the amount of dark energy \citep{2017A&A...604A..71H,2017PhRvD..96j3525M,2017MNRAS.469..394H,2018A&A...614A..13S,2020PASJ...72...26M}. Additionally, one can study the spatial distribution of the tSZ signal through two-point and higher-order correlation functions, as performed by the {\it Planck}, the Atacama Cosmology Telescope (ACT), and the South Pole Telescope (SPT) collaborations \citep{2016A&A...594A..22P,2020JCAP...12..045C,2015ApJS..216...27B}. The joint analysis of $n$-point statistics provides a powerful probe of large-scale structure formation across cosmic time \citep[e.g.,][]{2016A&A...594A..22P}.

The kSZ effect, on the other hand, is the result of CMB photons scattering off electrons that have bulk motion with respect to the CMB rest frame. As such, it is sensitive to both the gas velocity and its density, making it a powerful and versatile probe. On one hand, using pairwise and bispectrum estimators \citep{Hand:2012ui, 2018arXiv181013423S}, one can isolate the large-scale velocity field, which encodes details about the growth history, dark energy, primordial non-Gaussianities and modified gravity \citep{2016PhRvD..94d3522A,Munchmeyer:2018eey,2019PhRvD.100h3522P}. Such measurements have been conducted by the SPT, ACT, and {\it Planck} teams  \citep{2016MNRAS.461.3172S,2017JCAP...03..008D,2018A&A...617A..48P,2021PhRvD.104d3502C}. On the other hand, the kSZ effect is a powerful probe of baryonic physics. For example, stacking analyses \citep{Schaan:2020, schaan:2016} have been used to better understand the astrophysical processes at play in clusters \citep{Amodeo:2020mmu}, while projected field estimators can constrain the cosmic baryon fraction \citep{Hill:2016dta, Ferraro:2016ymw, Kusiak:2021hai, Bolliet:2022pze}. 
Thanks to its linear dependence on the gas density, the kSZ effect is also an excellent probe of the gas distribution in lower density environments, such as lower mass halos or the outskirts of larger halos, which are inaccessible with other techniques. As such, it is an excellent tool for measuring and calibrating baryon effects in weak lensing, one of the dominant sources of uncertainty on small scales for current and future weak lensing surveys, as first demonstrated in \citet{Amodeo:2020mmu}.

Finally, through the combination of tSZ and kSZ measurements, one can infer the temperature profiles of halos \citep{Battaglia:2017neq} and thus elucidate the thermodynamic processes that govern the ionized baryons in galaxies and galaxy clusters, including the hydrodynamics of the gas, star formation, and active galactic nuclei (AGN) activity \citep{Battaglia:2017neq,2018ApJ...865..109S,2021A&A...645A.112T,2021PhRvD.104d3503V}. Understanding group and cluster thermodynamics is essential to gaining insight into galaxy formation and evolution, and pinpointing the effect of baryons on the underlying dark matter, which can have substantial impact on cosmological analyses.

The upcoming measurements of the CMB temperature fluctuations with Advanced ACT, Simons Observatory and CMB-S4 \citep{2019JCAP...02..056A,2016arXiv161002743A} will lead to a significant increase in our constraining power, shifting the nature of our limitations from statistical uncertainties to theoretical modelling ones. To reliably extract the rich cosmological content of the secondary CMB anisotropies, we would need to overcome several obstacles related to the connection between gas physics and underlying matter as well as the distribution of electron pressure and density in the Universe. A possible venue for addressing these issues is provided by state-of-the-art hydrodynamical simulations, which only recently have reached sufficiently large volumes to be useful to cosmological analysis. These simulations paint a plausible picture of the real Universe and allow us to study the relations between various astrophysical quantities that cannot be disentangled from each other through (current) observations. In particular, group and cluster properties depend on multi-scale physical processes, such as AGN feedback and plasma physics \citep{2005Natur.433..604D,2016ApJ...824...79A, Kannan2016, Kannan2017, Barnes2019, Moser:2021llm}, so the only viable way of studying these processes in detail is through high-resolution, high-fidelity hydrodynamical simulations. Conversely, at scales that are not well-resolved, hydrodynamical simulations implement various semi-analytic prescriptions about complex and often poorly understood astrophysical processes such as supernova and AGN feedback, known as ``subgrid models.'' Since these simulations are not tuned to match the baryonic properties we infer from SZ measurements, comparisons with observations provide non-trivial tests for the subgrid physics models as implemented into these codes.

In this work, we utilize the largest-to-date hydrodynamical simulation, MTNG740 \citep{2022arXiv221010060P}, part of the MillenniumTNG suite of $N$-body and hydrodynamic simulations, to explore the properties of the electron pressure and density.
\teal{This paper rests on the shoulders of a large collection of previous works that have addressed scientific questions pertinent to SZ science via hydrodynamical simulations \citep[e.g.,][]{2001ApJ...549..681S,2002ApJ...579...16W,2002PhRvD..66d3002R,2010ApJ...725...91B,2016JCAP...08..058B,2016MNRAS.463.1797D}. Novel in this work is the detailed study of observed SZ quantities in juxtaposition to intrinsic ones, which have been the usual focus in the literature. In addition, we address questions related to the sensitivity of different probes (SZ, X-rays, FRBs) to halo properties such as optical depth, commenting on the contributions from intervening structure to the measured signal.} Capturing and understanding the electron and pressure properties on a per halo basis provides us with all the necessary ingredients for predicting the tSZ and kSZ contributions from isolated halos at arbitrary scales. However, halos in the real Universe are highly clustered, so in order to model correctly the observables from CMB experiments, it is essential to extend studies of the baryonic profiles beyond the one-halo term (i.e., the case of a halo
in isolation). Here, we model both the one- and two-halo terms (as well as the random uncorrelated signal) as a function of halo mass and comment on their respective fractional contributions to the integrated SZ quantities relevant to observations. Since the thermodynamic properties of the circumgalactic medium (CGM) and intergalactic medium (IGM) encode the effects of the halo assembly history, we additionally study the response of the pressure and density profiles to concentration and environment, two-halo properties that are well-known to play a key role in shaping galaxy and halo formation. 


The outline of this paper is as follows. In Section~\ref{sec:methods}, we describe the simulations and the SZ quantities that we measure. In Section~\ref{sec:results}, we study the various contributions to the SZ signal. We then provide fits to the SZ scaling relations (i.e., dependence on mass) across a wide range of halo masses and redshifts and explore the response of these relations to halo properties beyond mass. We quantify the effect of the assumed subgrid physics model on the SZ observables by making use of the smaller CAMELS simulations with extreme feedback variations. Finally, we conduct measurements of the SZ signal around red massive galaxies, mimicking the stacking analysis performed by jointly analyzing LSS and CMB experiments, and comment on the effect of `miscentering.' In Section~\ref{sec:conc}, we summarize our main findings and comment on their implications for SZ science.


\section{Methods}
\label{sec:methods}
\subsection{Simulations}

\subsubsection{MillenniumTNG}
\label{sec:mtng}

The simulation suite of the MillenniumTNG project consists of several hydrodynamical and $N$-body simulations of varying resolutions and box sizes, including also some simulations with a massive neutrino component. A detailed description of the full simulation set is given in \citet{2022arXiv221010060P,2022arXiv221010059H,2022arXiv221010068H,2022arXiv221010072H,2022arXiv221010065B,2022arXiv221010419B,2022arXiv221010066K,2022arXiv221010075C,Delgado2023,Ferlito2023}.

In this study, we employ the largest available full-physics simulation box and its dark matter only counterpart, containing $2 \times 4320^3$ and $4320^3$ resolution elements, respectively, in a comoving volume of $(500\,h^{-1} {\rm Mpc})^3$. These simulations use the same cosmological model as IllustrisTNG \citep{2017MNRAS.465.3291W,2018MNRAS.473.4077P,2018MNRAS.475..648P,2018MNRAS.475..624N,2018MNRAS.477.1206N,2018MNRAS.480.5113M,2018MNRAS.475..676S, 2019MNRAS.490.3234N,2019ComAC...6....2N,2019MNRAS.490.3196P}, and their resolution is comparable but slightly lower than that of the largest IllustrisTNG box, TNG300-1, with $2.1 \times 10^7 \hMsun$ for the baryons and $1.1 \times 10^8 \hMsun$ for the dark matter. In analogy with the naming conventions of IllustrisTNG, we refer to the hydrodynamic simulation as MTNG740 due to its boxsize of $L=500\,h^{-1}{\rm Mpc} = 738.12 \,{\rm Mpc}$, while for the dark matter only run we use MTNG740-DM. We note that 
\citet{2022arXiv221010060P} show that the galaxy properties predicted by MTNG740 are generally remarkably consistent with those of TNG300, in some properties even with TNG100. To first order, MTNG740 can thus be viewed as extending the IllustrisTNG model to a volume nearly 15 times larger while otherwise being very similar.

Halos (groups) are identified using the Friends-of-Friends (FoF) algorithm. Throughout the paper, as mass proxy, we adopt $M_{\rm 200c}$ defined as the mass of all the particles (dark matter, gas, stars and black holes) contained within $R_{\rm 200c}$ of the halo center, with $R_{\rm 200c}$ being the radius that encloses an overdensity $\Delta = 200$ with respect to the critical density of the Universe. The halo center is chosen as the location of the minimum gravitational potential within the FoF group.

\subsubsection{CAMELS}
\label{sec:camels}

To test the dependence of the thermal and kinematic Sunyaev-Zel'dovich measurements on the subgrid model adopted in the MTNG simulation, we also utilize a handful of the `extreme-feedback' (\texttt{EX}) boxes produced as part of the CAMELS project \citep{2021ApJ...915...71V,2023ApJS..265...54V}.

The CAMELS project offers a large set ($\sim$6000) of small-volume boxes ($L = 25\,h^{-1}{\rm Mpc}$) run with a variety of numerical simulation codes: 3 hydrodynamical ones, including the AREPO code used in generating the IllustrisTNG and MillenniumTNG suites, as well as an $N$-body (dark-matter-only) one. The parameter space spanned by the boxes covers two cosmological parameters ($\Omega_{\rm m}$ and $\sigma_8$) and four parameters corresponding to feedback from supernovae (SN1, SN2) and active galactic nuclei (AGN1, AGN2). The SN parameters encode the subgrid prescription for galactic winds, while the AGN parameters describe the efficiency of black hole feedback.

In this work, we employ the CAMELS \texttt{EX} suite composed of four CAMELS-TNG boxes that share the same cosmology ($\Omega_{\rm m} = 0.3$, $\sigma_8=0.8$, $\Omega_{\rm b} = 0.049$, $h = 0.6711$ and $n_s = 0.9624$), but differ in their astrophysical feedback. \texttt{EX\_0} has the fiducial feedback parameter values, while \texttt{EX\_1-3} represent the three extreme cases: 1) very efficient supernova feedback, 2) very efficient AGN feedback, and 3) no feedback. All simulations share the same initial conditions.

\subsection{SZ quantities}
\label{sec:sz_def}

The goal of this paper is to inform upcoming observations of the tSZ and kSZ signals via next-stage CMB and large-scale structure experiments. Therefore, the relevant quantities to explore are the integrated Compton-y parameter, $Y$, proportional to the size of the tSZ-induced temperature fluctuation, and the optical depth, $\tau$, proportional to the kSZ signal. We note that in reality, the kSZ signal measures the line-of-sight \textbf{velocity-weighted} optical depth, which differs non-trivially from the true optical depth of a cluster. Investigating the relationship between the two is one of the goals of this paper.

We denote the three quantities of interest as $Y_{{\rm 200c}, A}$, $\tau_{{\rm 200c}, A}$, $\tau_{{\rm kSZ, 200c}, A}$, where $A = \{{\rm sph}, \ {\rm cyl}\}$ stands for the signal enclosed in \textbf{either} a sphere centered around the halo/galaxy of interest with radius $R_{\rm 200c}$ \textbf{or} a cylinder passing through that halo/galaxy with radius $R_{\rm 200c}$ and length $L_{\rm box}$. We can calculate the first two from the simulations as follows:
\begin{equation}
Y_{{\rm 200c}, A} = \frac{\sigma_T}{m_e c^2} \int_{V_{{\rm 200c}, A}} P_e(\mathbf{r}) \ {\rm d}V,  
\end{equation}
\begin{equation}
\tau_{{\rm 200c}, A} = \sigma_T \int_{V_{{\rm 200c}, A}} n_e(\mathbf{r}) \ {\rm d}V , 
\end{equation}
where $\sigma_T$ is the Thomson cross section, $m_e$ is the electron mass, $c$ is the speed of light, $n_e(\mathbf{r})$ is the electron number density, and $P_e(\mathbf{r})$ is the electron pressure:
\begin{equation}
    P_e(\mathbf{r}) = n_e(\mathbf{r}) k_B T_e(\mathbf{r}) ,
\end{equation}
expressed in terms of the electron temperature $T_e$ and the Boltzmann constant $k_B$. The volume integral is performed either over a sphere, $V_{{\rm 200c, sph}}$, or a cylinder through the box, $V_{{\rm 200c, cyl}}$, along the line-of-sight.
In the case of the kSZ effect, the relevant quantity is
\begin{equation}
    b_{{\rm 200c}, A} = \sigma_T \int_{V_{{\rm 200c}, A}} n_e(\mathbf{r}) v_r(\mathbf{r}) \ {\rm d}V  ,
\end{equation}
where $v_r(\mathbf{r})$ is the radial (i.e. along the line-of-sight) velocity of the electrons within the integrated volume, which is dominated by the bulk velocity of the cluster. Assuming we have access to the cluster velocity (through e.g., velocity reconstruction), we can convert this quantity into optical depth by applying an effective inverse-noise weighting to each object \citep{Schaan:2020}:
\begin{equation}
\label{eq:tau_kSZ}
    \tau_{{\rm kSZ, 200c}, A} \equiv \frac{b_{{\rm 200c}, A} \ v_{r, \rm bulk}}{v_{\rm rms}^2},
\end{equation}
where $v_{r, \rm bulk}$ is the bulk velocity along the line of sight and $v_{\rm rms}$ is the root mean square (RMS) velocity of the clusters above the mass threshold of interest, $M_{\rm 200c} \gtrsim 10^{12}\, {\rm M}_\odot$. \teal{Bulk velocity here is defined as the mass-weighted velocity of all particles in the FoF group.} In principle, one can compute these quantities at any aperture, but in this work, we focus on $R_{\rm 200c}$ as a reasonable proxy of the virial radius (within 50\%), 
which is the typical scale of interest for SZ measurements. As a case study, we also consider one example where we vary the size of the aperture. 
Throughout the paper, we utilize these quantities in units of ${\rm Mpc^2}$.

We note that the `sph' quantities (measured in a sphere around the halo/galaxy of interest) are not typically observable and can be thought of as `intrinsic' to the object. On the other hand, in real observations, the measured signal around a given object receives contributions from intervening structure along the line-of-sight. Thus, our `cyl' quantities (measured in a cylinder through the entire box) can be thought of as `observable' and studying how the two differ can be invaluable to interpreting observations. Note that the optical depth quantity $\tau_{\rm 200c, cyl}$ is not measurable through the SZ effect, but can be accessed, for example, in Fast Radio Bursts (FRBs) observations, or through CMB measurements of ``patchy-$\tau$'' \citep{Dvorkin:2008tf}.



\section{Results}
\label{sec:results}

In this study, we focus on interpreting the measurements of SZ quantities around halos identified in the hydrodynamical simulation MTNG740 (see Section~\ref{sec:mtng}). In particular, we utilize halo samples extracted at 4 distinct simulation snapshots: $z = 0, \ 0.25, \ 0.5, \ {\rm and} \ 1$, focusing on the most massive objects above a threshold of $10^{12} \, {\rm M}_\odot$, which corresponds to $\gtrsim$550,000 objects per snapshot. Most of the plots are shown for $z = 0.5$, where we expect the sensitivity of current and near-future observations to peak, but we make quantitative and qualitative statements about other redshifts throughout the text. As a case study, in Section \ref{sec:DESILRG}, we explore measurements around stellar-mass-selected galaxies, akin to a luminous red galaxy (LRG) sample, meant to mimic a joint study between a modern large-scale structure survey such as CMASS and DESI and a high-resolution CMB experiment.

\subsection{Observed and intrinsic signal}
\label{sec:rand}

Of major interest to the analysis of SZ observations is determining the various sources that contribute to the signal and interpreting their astrophysical significance. The SZ profile of a cluster, as measured on the CMB map, receives contributions from the cluster of interest (i.e., `one-halo term'), nearby clusters with a correlated spatial distribution (i.e., `two-halo term'), and uncorrelated (`random') structure along the line-of-sight.
To understand the respective size of each of these terms, we explore the mean relations between the SZ quantities, defined in Section \ref{sec:methods}, as a function of halo mass. In particular, we compare three modes of the measurement: `intrinsic' (spherical), `observed' (cylindrical), and `randoms-subtracted observed' (cylindrical). The subtraction of randoms is performed by adopting a Compensated Aperture Photometry (CAP) filter, which is typically used in observational analyses to reduce the contributions from the primary (degree-size) CMB  fluctuations. It is obtained by summing the signal in a disk of certain radius (in our case $R_{200c}$) centered on an object of interest and subtracting a larger concentric ring (with inner radius\footnote{The larger inner radius of the ring is to avoid subtracting a meaningful part of the signal. The purpose of the ring is to subtract the random contribution.} $4 R_{200c}$) around that same object of the same area. 

\begin{figure}
    \includegraphics[width=0.48\textwidth]{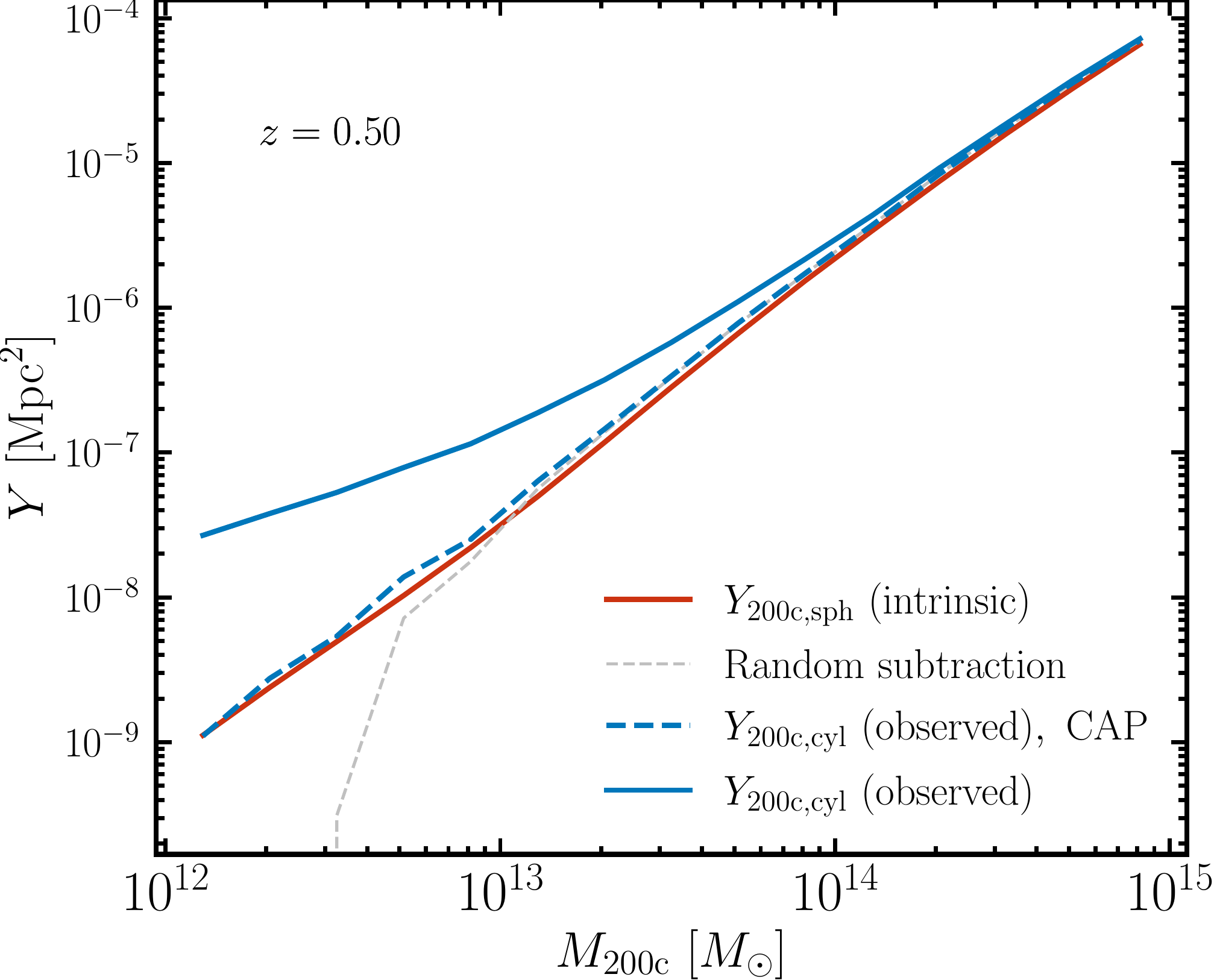}
    \caption{Mean scaling relation between the integrated Compton-y parameter, $Y$, and halo mass, $M_{\rm 200c}$, at $z = 0.5$. For the lowest-mass halos, the intrinsic $Y$ (i.e. electron pressure enclosed in a sphere of radius $R_{\rm 200c}$ around the halo center) differs by an order of magnitude from the observed $Y$ (computed by integrating the signal in a cylinder through the box) due to the contribution of random uncorrelated structures along the line of sight. Luckily, we largely eliminate this spurious effect when we adopt a Compensated Aperture Photometry (CAP) filter, and are left with the combined contribution of the one- and two-halo term (i.e. the intrinsic signal and the signal coming from nearby correlated structure). Alternatively, we can subtract the random contribution by measuring the SZ effect in random parts of the sky (silver dashed curve) and subtracting that from the signal. While this yields an excellent agreement with the CAP approach for intermediate and massive halos, we see that it overpredicts the random contribution for small-mass halos, as the selection of low-mass halos preferentially selects regions of lower density and thus, weaker SZ.}
    \label{fig:Y_M200c}
\end{figure}

In Fig.~\ref{fig:Y_M200c}, we show the mean relation between the integrated Compton-y parameter and halo mass at $z = 0.5$. Here we have split the data into 15 mass bins ranging between $M_{\rm 200c} > 10^{12} \, {\rm M}_\odot$ and $M_{\rm 200c} < 10^{15} \, {\rm M}_\odot$ and compute the mean signal in each mass bin for the three measurement modes. The slope of the intrinsic relation curve (shown in red) exhibits a slight mass-dependence, i.e., a smooth break in the power law, which we address in subsequent sections. The `observed' curve without CAP filtering (shown in solid blue) looks noticeably different from the intrinsic one, especially in the low-mass regime, where the measured $Y$ signal for a given halo is on average much larger than the underlying signal. This suggests that the random contribution is quite large for halos with masses below $\log(M_{\rm 200c}) < 13.5$, as expected, due to their lower temperature and electron content, leading to a smaller intrinsic $Y$. On the other hand, the CAP-filtered observed curve (dashed blue) demonstrates that the CAP filtering does an excellent job of removing the spurious uncorrelated contributions along the line-of-sight. We additionally test a version of the randoms subtraction procedure in which we subtract the background contribution by measuring the tSZ signal from disks carved in random locations on the map. This is shown as a faint gray line on the figure. We note that measuring the randoms in that manner leads to much more noisy results,
as the signal varies greatly along each line-of-sight. Additionally, in purposefully selecting empty regions in the sky and low-mass halos, we are also implicitly introducing a bias. 
Once the CAP filter has been applied, we are left with a 1-halo and a 2-halo term contribution. We leave the detailed modeling of the latter for a subsequent analysis. From hereon, when handling SZ observables, we always show the CAP-filtered quantities.

\begin{figure}
    \includegraphics[width=0.48\textwidth]{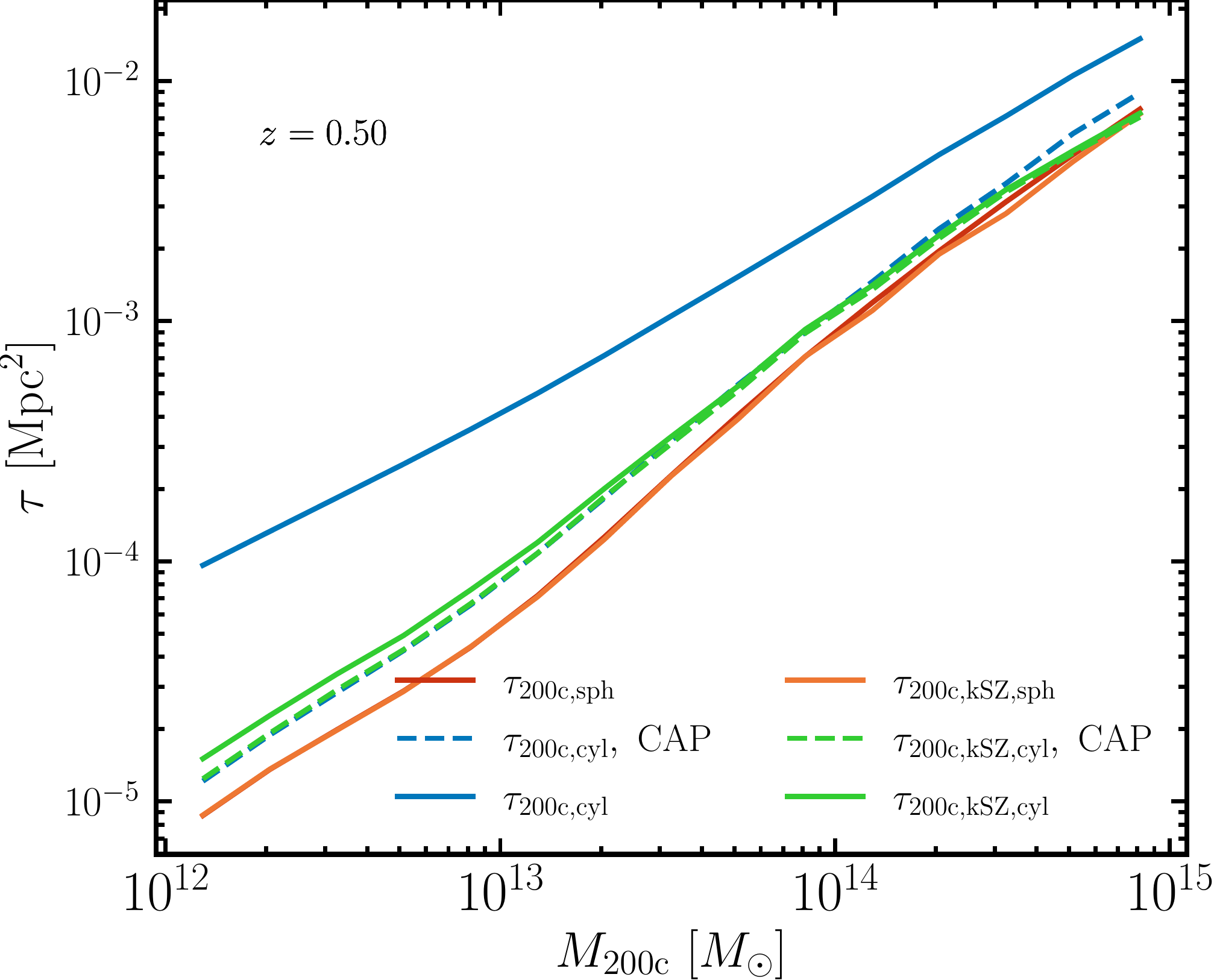}
    \caption{Mean scaling relation between the optical depth, $\tau$, defined in several different ways, and halo mass, $M_{\rm 200c}$, at redshift $z = 0.5$. The intrinsic $\tau$ (i.e. optical depth enclosed in a sphere around the halo center) is most akin to the X-ray-inferred optical depth of a cluster (since the X-ray signal scales as $n_e^2$ and has a weak dependence on temperature). The cylindrical $\tau$ is measured in e.g., FRB analysis, and $\tau_{\rm kSZ, cyl}$ is what we infer from kSZ measurements. Due to the decorrelation of velocities along the line of sight, $\tau_{\rm kSZ}$ is localized to a cylinder around the halo of interest of length equal to several tens of Mpc and is thus mostly made up of the 1-halo and 2-halo combined signal. At high halo masses, the signal is dominated by the halo of interest and almost insensitive to the halo surroundings.}
    \label{fig:tau_M200c}
\end{figure}

A long-standing issue in interpreting the kSZ signal is understanding how the optical depth inferred from the measurements compares with the integrated optical depth and the intrinsic optical depth of a cluster. We address this question in Fig.~\ref{fig:tau_M200c}, which shows the mean relation between optical depth, defined in several different ways, and halo mass. Roughly speaking, $\tau_{\rm kSZ, 200c, cyl}$ is what is inferred from kSZ observations; $\tau_{\rm 200c, cyl}$ is the optical depth of FRBs or patchy-$\tau$; $\tau_{\rm 200c, sph}$ is similar to the X-ray inferred $\tau$, as the X-ray signal scales as $n_e^2$ times the temperature to a weak power. 

As before, on average, we see that the intrinsic measurement of optical depth (red solid) roughly follows a power law with a break around $10^{14} \, {\rm M}_\odot$, corresponding to the regime where self-similarity arguments break down due to feedback processes, as discussed next. The kSZ-inferred intrinsic optical depth (orange solid) is almost indistinguishable from the spherical $\tau$ (red curve), suggesting that the assumption that the kSZ profile is well-approximated by the product between optical depth and bulk velocity holds true to a high degree of accuracy, indicating that the thermal velocities of the electrons inside the halo projected along the line-of-sight cancel, as expected. It is striking to note that the cylindrical measurement of $\tau$ (solid blue) receives a large contribution from random substructures along the line-of-sight, which add almost an order of magnitude to the intrinsic signal for the lower-mass halos we consider. This is unlike the case of the integrated Compton-y parameter, which is additionally weighted by temperature and hence receives its bulk signal from denser structures. As before, we see that the $\tau$ we are left with (dashed blue) after applying the CAP filter successfully removes the random contribution and leaves us with the combined one- and two-halo term. 

One might naively expect that the cylindrical kSZ-inferred optical depth (solid green) should yield a similar result to the observed optical depth (blue solid). In reality, that curve is much more similar to the CAP-filtered optical depth (dashed blue), as the only non-vanishing contribution to the halo-velocity-weighted kSZ signal is from structures moving at coherent velocities. For smaller-mass halos, which tend to be infalling towards larger clusters, we expect that much of the surrounding substructure will have similarly correlated velocities (i.e., pointing towards a larger cluster), which is indeed what we observe in the figure. Once we remove the random contribution via the CAP filter, the kSZ-inferred optical depth (dashed green) becomes almost identical to the dashed blue curve, and that holds true all the way up to the most massive clusters. These objects tend to act as gravitational attractors to the structures surrounding them, leading to symmetric line-of-sight velocities with respect to their centers that cancel in the kSZ measurement. Thus, for these massive clusters the kSZ-inferred optical depth approximates their true optical depth. Another way of thinking about this phenomenon is by studying the large-scale velocity fields: one can show that radial velocities decorrelate much faster along the line of sight (i.e., they have a shorter correlation length) than the matter density does; thus, assuming that the decorrelation length is kept fixed and centered at the halo of interest, as we observe less massive and thus smaller halos, we will encompass more (relative to larger halos) of the two-halo term and the halo surroundings, and vice versa for massive clusters, leading to the observed behavior across the different mass scales.

\subsection{Fits to the scaling relations}
\label{sec:fits}

\subsubsection{Expected scaling with mass}
\label{sec:scaling}

For massive halos, where the assumption that the dominant source of energy
input into the intra-cluster 
medium is gravitational, one can derive scaling relations that connect intrinsic halo properties such as its mass to observed halo properties such as its measured SZ or X-ray signal. 
Under the assumption of a collapsed isothermal sphere, the infalling gas into a given halo eventually reaches a temperature approximately equal to the virial temperature of the halo, which can be related to the halo potential, $\Phi$, via the virial theorem:
\begin{equation}
k_B T_\Delta \propto -\frac{1}{2}\Phi = \frac{G M_\Delta \mu \, m_p}{2 r_\Delta},
\end{equation}
where $k_B$ is the Boltzmann constant, $m_p$ is the proton mass, $\Delta=200$ for our case, and $\mu$ is the mean molecular weight. Since halo mass scales with its size as:
\begin{equation}
M_\Delta \propto \Delta \rho_{\rm crit}(z) r_\Delta^3 \propto E^2(z) r_\Delta^3,
\end{equation}
the self-similar scaling relation between temperature and total halo mass is:
\begin{equation}
T_\Delta \propto \frac{M_\Delta}{r_\Delta} \propto M_\Delta^{2/3} E^{2/3}(z).
\end{equation}

In the case of the thermal Sunyaev-Zel'dovich effect, the integrated Compton-y parameter is a measure of the total thermal energy of the ICM and is proportional to $Y \propto \int n_e T_e {\rm d}V \propto M_{\rm gas} \bar T_e $, where $n_e$ and $T_e$ are the electron number density and temperature, and $\bar T_e$ is the mass-weighted mean temperature. The baryonic content of halos is dominated by the hot ICM, and thus the gas fraction inside halos is roughly equal to the mass fraction of baryons $f_g \approx \frac{M_{\rm baryons}}{M_{\rm total}}$, and the baryon fraction in the Universe, $ \frac{\Omega_b}{\Omega_m}$, implying $Y \propto M_{\rm total}^{5/3}$. Similarly, $\tau_{\rm kSZ} \propto \int n_e {\rm d}V \propto M_{\rm gas} \propto M_{\rm total}$, which shows that the signal is sensitive to a different scaling of the baryonic quantities. Provided one can recover the velocity via, e.g., reconstruction, the joint utilization of tSZ and kSZ measurements can be highly potent for constraining the thermodynamic properties of galaxies, groups and clusters.

However, at lower halo masses, 
\teal{as feedback energy gets injected into the group or cluster, the gas entropy increases above the self-similar value, causing a slow outflow that reduces the baryon fraction in the clusters.}
This results in non-trivial deviations from the self-similar relations derived above, which manifest themselves as ``breaks'' in the assumed power law. \teal{This phenomenon has been observed in both simulations \citep[e.g.,][]{2015MNRAS.451.3868L,2020MNRAS.498.3061R,2022MNRAS.516.4084Y,2022arXiv220511537P,2022arXiv220511528P,2023PNAS..12002074W} and observations \citep[e.g.,][]{2015ApJ...808..151G,2018MNRAS.475..532O,2018PhRvD..97h3501H,2021MNRAS.501.2467S,2022PhRvD.105l3526P}}, and can be addressed by introducing a mass-dependence to the slope of the power law. Below, we adopt the `smoothly broken power law' formalism, proposed in \citet{2022arXiv220511528P}, aimed at capturing the mass-dependent behavior of the scaling relations.

\subsubsection{Smoothly broken power law}
\label{sec:sbpl}

\begin{figure*}
    \includegraphics[width=0.48\textwidth]{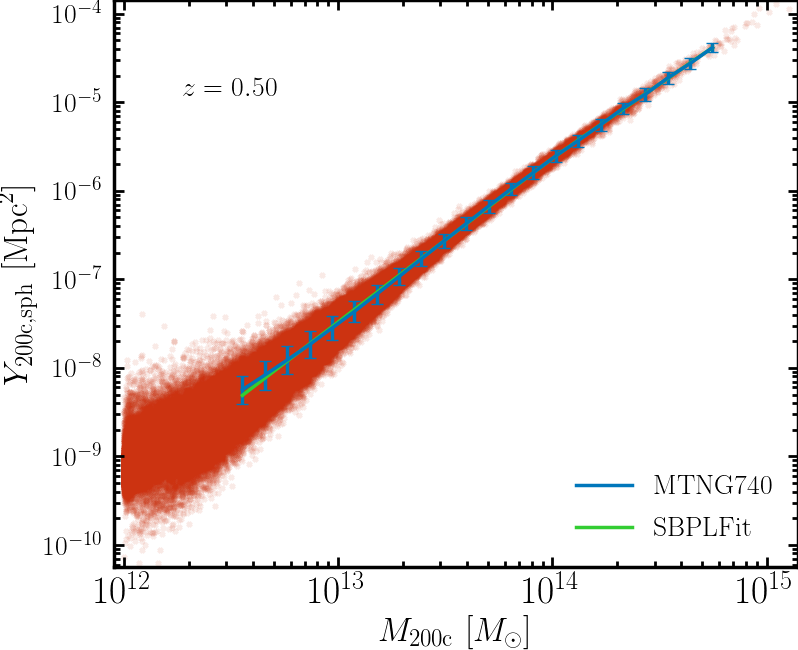}
    \includegraphics[width=0.48\textwidth]{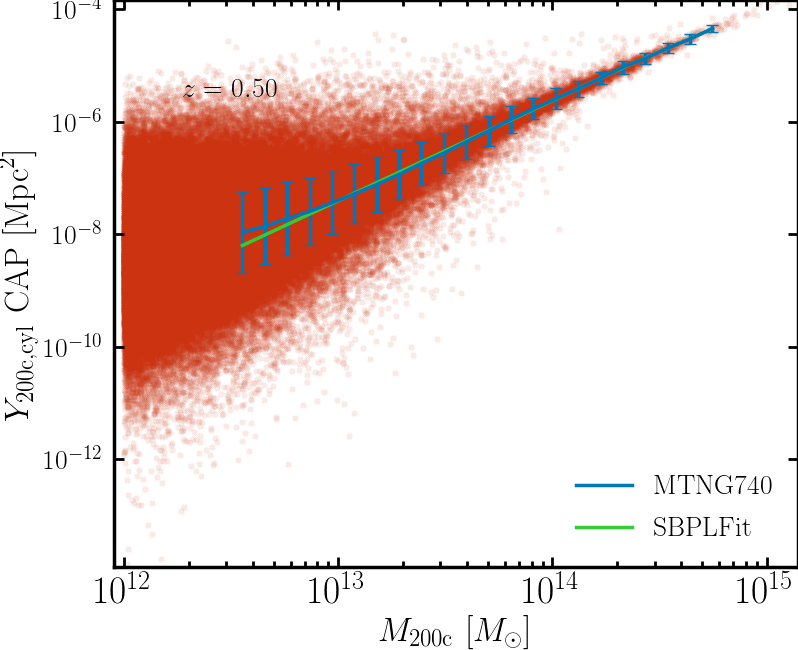}\\
    \includegraphics[width=0.48\textwidth]{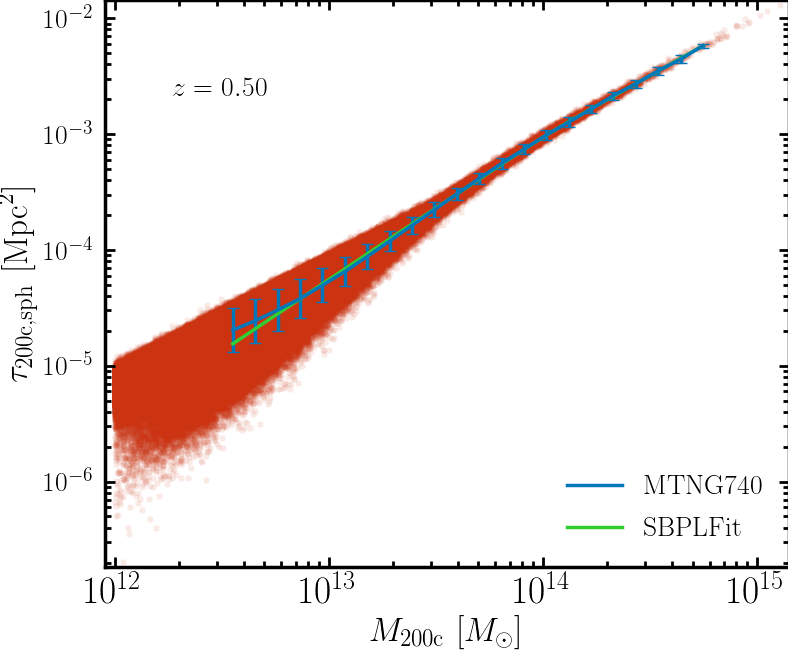}
    \includegraphics[width=0.48\textwidth]{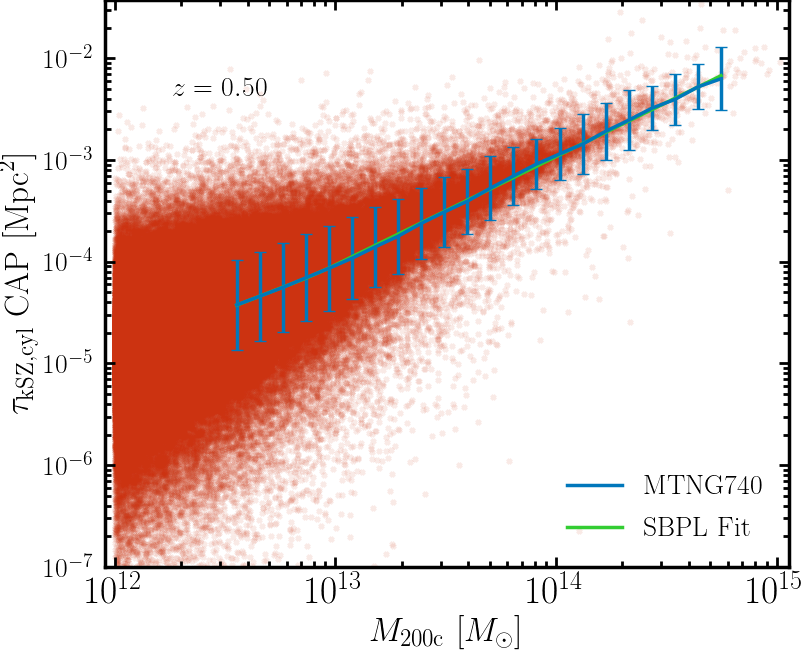}
    \caption{Smoothly broken power law (SBPL) fits for the $Y-M$ and $\tau-M$ relations (see Section \ref{sec:sbpl}). The left panels show the scatter plot for the intrinsic quantities, whereas the right panels for the observed ones, as measured in a cylinder through the box with a CAP filter. The data curves with error bars are computed using the geometric mean and standard deviation in log-space. We see clear deviations from self-similarity on the left for halos below $M_{\rm 200c} \lesssim 10^{14}\, {\rm M}_\odot$, for which feedback processes dominate the baryon distribution within the halo. This break is somewhat obscured on the right where structures along the line of sight add extra noise. 
    \teal{Note that here we define the kSZ-inferred optical depth as the kSZ signal divided by the line-of-sight velocity rather than by applying the inverse-variance weighting used in the rest of the paper (see discussion in Section~\ref{sec:sbpl}.}
    The best-fit parameters are shown in Table~\ref{tab:SBPL}.}
    \label{fig:SBPL}
\end{figure*}

\begin{table*}
\begin{center}
\begin{tabular}{c | c c c c c | c}
 \hline\hline
 Scaling & $A$ & $\alpha_1$ & $\alpha_2$ & $\delta$ & $\log_{10}(X_{\rm pivot})$ & Redshift \\ [0.5ex]
 \hline
$Y_{\rm 200c, sph} - M_{\rm 200c}$ & -5.692$\pm{0.001}$ & 2.003$\pm{0.001}$ & 1.649$\pm{0.014}$ & 0.212$\pm{0.033}$ & 13.912$\pm{0.020}$ & 0.00  \\ [1ex]
$Y_{\rm 200c, cyl} \ {\rm CAP} - M_{\rm 200c}$ & -5.684$\pm{0.003}$ & 1.826$\pm{0.006}$ & 1.652$\pm{0.037}$ & 0.016$\pm{0.065}$ & 14.224$\pm{0.056}$ & 0.00  \\ [1ex]
$\tau_{\rm 200c, sph} - M_{\rm 200c}$ & -3.035$\pm{0.001}$ & 1.396$\pm{0.002}$ & 1.037$\pm{0.013}$ & 0.216$\pm{0.027}$ & 13.959$\pm{0.015}$ & 0.00  \\ [1ex]
$\tau_{\rm kSZ, 200c} \ {\rm CAP} - M_{\rm 200c}$ & -2.957$\pm{0.002}$ & 1.158$\pm{0.005}$ & 0.805$\pm{0.253}$ & 0.324$\pm{0.211}$ & 14.432$\pm{0.288}$ & 0.00  \\ [1ex]
$Y_{\rm 200c, sph} - \tau_{\rm 200c, sph}$ & -5.642$\pm{0.000}$ & 1.465$\pm{0.000}$ & 1.587$\pm{0.025}$ & 0.075$\pm{0.080}$ & -2.894$\pm{0.047}$ & 0.00  \\ [1ex]
$Y_{\rm 200c, cyl} \ {\rm CAP} - \tau_{\rm kSZ, 200c} \ {\rm CAP}$ & -5.946$\pm{0.005}$ & 1.627$\pm{0.010}$ & 1.590$\pm{0.018}$ & 0.010$\pm{0.000}$ & -3.449$\pm{0.015}$ & 0.00  \\ [1ex]
\hline
$Y_{\rm 200c, sph} - M_{\rm 200c}$ & -5.666$\pm{0.001}$ & 1.913$\pm{0.001}$ & 1.635$\pm{0.015}$ & 0.216$\pm{0.031}$ & 14.004$\pm{0.024}$ & 0.25  \\ [1ex]
$Y_{\rm 200c, cyl} \ {\rm CAP} - M_{\rm 200c}$ & -5.662$\pm{0.006}$ & 1.783$\pm{0.016}$ & 1.533$\pm{0.035}$ & 0.185$\pm{0.232}$ & 14.500$\pm{0.000}$ & 0.25  \\ [1ex]
$\tau_{\rm 200c, sph} - M_{\rm 200c}$ & -3.031$\pm{0.001}$ & 1.303$\pm{0.001}$ & 1.022$\pm{0.011}$ & 0.211$\pm{0.028}$ & 14.025$\pm{0.017}$ & 0.25  \\ [1ex]
$\tau_{\rm kSZ, 200c} \ {\rm CAP} - M_{\rm 200c}$ & -2.963$\pm{0.013}$ & 1.087$\pm{0.206}$ & 0.865$\pm{0.224}$ & 0.010$\pm{0.343}$ & 14.500$\pm{1.498}$ & 0.25  \\ [1ex]
$Y_{\rm 200c, sph} - \tau_{\rm 200c, sph}$ & -5.616$\pm{0.001}$ & 1.716$\pm{0.008}$ & 1.505$\pm{0.001}$ & 0.056$\pm{0.015}$ & -4.093$\pm{0.013}$ & 0.25  \\ [1ex]
$Y_{\rm 200c, cyl} \ {\rm CAP} - \tau_{\rm kSZ, 200c} \ {\rm CAP}$ & -5.990$\pm{0.004}$ & 1.338$\pm{0.130}$ & 1.631$\pm{0.019}$ & 0.308$\pm{0.242}$ & -4.093$\pm{0.174}$ & 0.25  \\ [1ex]
\hline
$Y_{\rm 200c, sph} - M_{\rm 200c}$ & -5.635$\pm{0.002}$ & 1.853$\pm{0.001}$ & 1.665$\pm{0.031}$ & 0.106$\pm{0.115}$ & 14.010$\pm{0.063}$ & 0.50  \\ [1ex]
$Y_{\rm 200c, cyl} \ {\rm CAP} - M_{\rm 200c}$ & -5.624$\pm{0.002}$ & 1.780$\pm{0.005}$ & 1.658$\pm{0.022}$ & 0.010$\pm{0.000}$ & 14.318$\pm{0.017}$ & 0.50  \\ [1ex]
$\tau_{\rm 200c, sph} - M_{\rm 200c}$ & -3.028$\pm{0.001}$ & 1.236$\pm{0.001}$ & 1.041$\pm{0.013}$ & 0.126$\pm{0.045}$ & 14.028$\pm{0.028}$ & 0.50  \\ [1ex]
$\tau_{\rm kSZ, 200c} \ {\rm CAP} - M_{\rm 200c}$ & -2.967$\pm{0.005}$ & 0.843$\pm{0.023}$ & 1.064$\pm{0.011}$ & 0.010$\pm{0.055}$ & 12.918$\pm{0.042}$ & 0.50  \\ [1ex]
$Y_{\rm 200c, sph} - \tau_{\rm 200c, sph}$ & -5.590$\pm{0.001}$ & 1.714$\pm{0.005}$ & 1.530$\pm{0.002}$ & 0.076$\pm{0.022}$ & -4.019$\pm{0.009}$ & 0.50  \\ [1ex]
$Y_{\rm 200c, cyl} \ {\rm CAP} - \tau_{\rm kSZ, 200c} \ {\rm CAP}$ & -6.026$\pm{0.004}$ & 1.214$\pm{0.100}$ & 1.623$\pm{0.009}$ & 0.187$\pm{0.063}$ & -4.079$\pm{0.075}$ & 0.50  \\ [1ex]
\hline
$Y_{\rm 200c, sph} - M_{\rm 200c}$ & -5.578$\pm{0.002}$ & 1.788$\pm{0.002}$ & 1.576$\pm{0.286}$ & 0.115$\pm{0.166}$ & 14.245$\pm{0.230}$ & 1.00  \\ [1ex]
$Y_{\rm 200c, cyl} \ {\rm CAP} - M_{\rm 200c}$ & -5.549$\pm{0.005}$ & 1.780$\pm{0.010}$ & 1.606$\pm{0.296}$ & 0.047$\pm{0.240}$ & 14.200$\pm{0.255}$ & 1.00  \\ [1ex]
$\tau_{\rm 200c, sph} - M_{\rm 200c}$ & -3.029$\pm{0.001}$ & 1.159$\pm{0.001}$ & 1.000$\pm{0.053}$ & 0.089$\pm{0.117}$ & 14.139$\pm{0.076}$ & 1.00  \\ [1ex]
$\tau_{\rm kSZ, 200c} \ {\rm CAP} - M_{\rm 200c}$ & -2.982$\pm{0.009}$ & 0.839$\pm{0.013}$ & 1.033$\pm{0.018}$ & 0.081$\pm{0.045}$ & 12.957$\pm{0.066}$ & 1.00  \\ [1ex]
$Y_{\rm 200c, sph} - \tau_{\rm 200c, sph}$ & -5.532$\pm{0.002}$ & 1.708$\pm{0.007}$ & 1.565$\pm{0.004}$ & 0.086$\pm{0.036}$ & -3.935$\pm{0.027}$ & 1.00  \\ [1ex]
$Y_{\rm 200c, cyl} \ {\rm CAP} - \tau_{\rm kSZ, 200c} \ {\rm CAP}$ & -6.047$\pm{0.008}$ & 1.094$\pm{0.197}$ & 1.709$\pm{0.050}$ & 0.361$\pm{0.255}$ & -3.977$\pm{0.120}$ & 1.00  \\ [1ex]
 \hline
 \hline
\end{tabular}
\end{center}
\caption{Best-fit values and jackknife errors for the smoothly broken power law (SBPL) model (see Section \ref{sec:sbpl}) at four distinct time epochs: $z = 0, \ 0.25, \ 0.5, \ {\rm and} \ 1$, fit to the observed (cylinder with a CAP filter) and intrinsic (sphere) SZ quantities, integrated Compton-y parameter ($Y$) and optical depth ($\tau$), measured in the MTNG740 hydrodynamical simulation.}
\label{tab:SBPL}
\end{table*}

At the crux of the `smoothly broken power law' (SBPL) model is the notion that the deviation from self-similarity should occur smoothly, as one transitions from high-mass to low-mass halos. Thus, one can adopt a smoothly varying slope to the power law, which asymptotically reproduces the behaviour of a simple power-law for the highest mass halos. The slope at a given mass (or more generally, $X$) is defined as:
\begin{equation}
\alpha_{\rm SBPL} = \frac{{\rm d} \log Y}{{\rm d} \log X} = \frac{\alpha_2 - \alpha_1}{2} \tanh \left[ \frac{1}{\delta} {\rm log}_{10} \left(\frac{X}{X_{\rm pivot}} \right) \right] + \frac{\alpha_2 + \alpha_1}{2}.\label{eqn:eqnSBPLslope}
\end{equation}
The smoothly broken power law thus has a total of five free parameters ($A_1$, $\alpha_1$, $\alpha_2$, $\delta$, $X_{\rm pivot}$):
\begin{equation}
\frac{Y(X)}{Y(X_{\rm norm})} =  \left( \frac{X}{X_{\rm norm}} \right)^{ \left(\frac{\alpha_2 + \alpha_1}{2}\right)} \left[ \frac{{\rm cosh} \left(\frac{1}{\delta} {\rm log}_{10} \left( \frac{X}{X_{\rm pivot}}\right) \right)  }{ {\rm cosh} \left( \frac{1}{\delta} {\rm log}_{10} \left(\frac{X_{\rm norm}}{X_{\rm pivot}}\right)\right) } \right]^{\left( \frac{\alpha_2 - \alpha_1}{2}\right) \, \delta \, {\rm ln} 10}.\label{eqn:eqnSBPL}
\end{equation}
Here, $X_{\rm pivot}$ is the pivot scale of the transition, $\delta$ is a measure of the width of the transition between the two slopes ($\alpha_1$ for $X \ll X_{\rm pivot}$ and $\alpha_2$ for $X \gg X_{\rm pivot}$, respectively).

One of the goals of our work is to provide accurate estimates of the slopes characterizing the underlying connection between different SZ quantities and halo mass. To this end, we perform fits to the simulation data using a weighted least-squares fitting routine that assigns different weights based on the level of scatter in each mass bin. In particular, we characterize the relations via the logarithm of the geometric mean, which has the benefit of preserving the original slope of the data in log-log space, and the scatter around the geometric mean in log-space.
We note that while using the geometric mean has the desirable property of describing the bulk behavior of the signal in a less biased fashion \citep[see][]{2022arXiv220511537P}, the arithmetic mean is the preferred quantity to use in analyzing observations, as in averaging the noisy signal from data, random contributions from, e.g., the primary CMB or systematics cancel. Were we to use geometric mean instead, we might obtain a non-trivial bias contribution. We derive uncertainties on our best-fit parameters by performing the same analysis on $100$ jackknife samples.

In Fig.~\ref{fig:SBPL} and Table~\ref{tab:SBPL}, we show fits to the intrinsic and observed quantities: $Y-M$, $\tau-M$ and $Y-\tau$, the latter of which is central to breaking the so-called `optical depth degeneracy,' which is the main obstacle to utilizing kSZ probes on their own for cosmological analyses \citep{2018arXiv181013423S}. It is evident from Fig.~\ref{fig:SBPL} that the scatter in the observed quantities (right panels) is much larger than in the intrinsic quantities (left panels), especially for lower mass clusters for which the relative contribution of structures along the line-of-sight is larger. In real data, it is not unusual to mask or throw away smaller groups whose angular coordinates are in close proximity to larger groups clusters, which would reduce some of the scatter. We additionally note that the reported best-fit values are somewhat dependent on the choice of binning; for the mass-based relations, we set the number of bins to 23 in the range of $\log M_{\rm 200c} = 12.5 - 14.8$ (for redshift $z = 1$, we stop at 14.6, as we do not have examples of halos above that mass).

The fits to the $Y-M$ relation shown in Table~\ref{tab:SBPL} reveal several interesting features. As noted in previous works \citep[e.g.,][]{2023PNAS..12002074W}, the intrinsic relation, $Y_{\rm 200c, sph} - M_{\rm 200c}$, obeys self-similarity, i.e. the slope approaches 5/3, past $\log M_{\rm 200c} \approx 14$. We note that if we were to choose a smaller aperture such as $R_{\rm 500c}$, the pivot point would move to lower halo masses (compare with \citealt{2022arXiv220511537P}, \teal{whose pivot mass is shifted to lower halo masses, $\log M \sim 13.5$ as a result of the aperture choice; in the high-mass regime, our findings are in excellent agreement}), as the AGN feedback processes dislodge the baryons from the central regions of the cluster more easily than from the outskirts, violating self-similarity. It is interesting to note that the break in the observed $Y_{\rm 200c, cyl} \ {\rm CAP} - M_{\rm 200c}$ relation is somewhat obscured by the presence of the two-halo term (see also Fig.~\ref{fig:Y_M200c}), which can be seen in noting that the power law indices $\alpha_1$ and $\alpha_2$ are more similar in the observed case compared with the intrinsic case. We can understand that by noting that the two-halo term has a relatively larger contribution for lower-mass halos than higher-mass halos, as indicated by Fig.~\ref{fig:Y_M200c}, \teal{which compensates for the slope difference at lower masses and leads to lower values of $\alpha_1$ compared with the intrinsic relations.}

Moving our attention to the $\tau-M$ fits, we notice a similar trend to $Y-M$ for the intrinsic signal, $\tau_{\rm 200c, sph} - M_{\rm 200c}$: it exhibits self-similar behavior, i.e. $\tau \propto M_{\rm 200c}$, for masses larger than $M_{\rm 200c} = 10^{14} \, {\rm M}_\odot$ and appears to be steeper for lower mass-halos for which AGN feedback is stronger than gravity. On the other hand, the observed relation, $\tau_{\rm kSZ, 200c} \ {\rm CAP} - M_{\rm 200c}$, also approaches unity for high-mass halos, but we see that the scatter is much larger.
\teal{We note that our definition of the kSZ-inferred optical depth, $\tau_{\rm kSZ, 200c} \ {\rm CAP}$ differs from the definition used in the rest of the paper: instead of applying inverse-variance weighting (i.e., Eq.~\ref{eq:tau_kSZ}), we directly divide the kSZ signal by the line-of-sight velocity, $v_{r, {\rm bulk}}$. We stress that when studying the mean behavior, Eq.~\ref{eq:tau_kSZ} offers the optimal estimator of the optical depth, as it takes care of outliers such as measurements around halos with small line-of-sight velocities and along lines-of-sight that receive large contributions from intervening structure yielding a sign difference between the kSZ signal and the velocity of the halo. Additionally, when using real data, the reconstructed velocity is noisy, motivating the use of inverse-variance weighting. When exploring the optical depth on an object-by-object basis, using Eq.~\ref{eq:tau_kSZ} picks up additional scatter from the ratio between the root mean square velocity and the line-of-sight velocity, which is the reason we opt for the alternative definition when providing fits to the scaling relations and displaying the scatter plots. In order to measure the geometric mean, we remove all instances of negative kSZ-inferred tau and lines-of-sight velocity smaller than 20\% of the root mean square velocity. Results of the fits are shown in Table~\ref{tab:SBPL}. As in the observed $Y-M$ relation, there is a tendency of flattening the fitted curve in the low-mass regime, although the fits show no clear preference for a break at $z = 0$ and 0.25.}

Finally, the $Y-\tau$ relationship fits are shown in Table~\ref{tab:SBPL} for the intrinsic $Y_{\rm 200c, sph} - \tau_{\rm 200c, sph}$ and observed $Y_{\rm 200c, cyl} \ {\rm CAP} - \tau_{\rm kSZ, 200c} \ {\rm CAP}$ signal. We note that the break in the power law is not as evident, being that both $Y$ and $\tau$ are influenced by the same astrophysical processes related to AGN feedback. As a result, the $\alpha_1$ and $\alpha_2$ values tend to be either very similar or measured with a larger error. In analyzing observations, having a measurement of both the kSZ and tSZ signal around a given group or cluster, one can make use of these relations to break the degeneracy between radial velocity and optical depth. Having a clean measurement of the infall velocity serves as a powerful probe in cosmology, allowing us to place constraints on the growth of structure, modified gravity and primordial non-Gaussianities \citep[see e.g.,][]{2016PhRvD..94d3522A,Munchmeyer:2018eey,2019PhRvD.100h3522P}. However, a caveat that we comment on in Section \ref{sec:camels_sub}, is that the resulting expressions are highly sensitive to the subgrid model of the simulation (i.e. implementation of feedback), which could introduce substantial systematic errors if not properly accounted for.

Finally, when handling real data, we often want to study the tSZ and kSZ profiles as a function of distance from the galaxy center.
Thus, an alternative format for reporting our findings would be as fits to a functional form such as the generalized Navarro-Frenk-White profile (GNFW), which describes the full shape of the electron pressure and density profiles, the two ingredients needed to predict the SZ observables. We leave this task for future investigation.

\subsection{Dependence on assembly bias}
\label{sec:ass_bias}

\begin{figure}
    \includegraphics[width=0.45\textwidth]{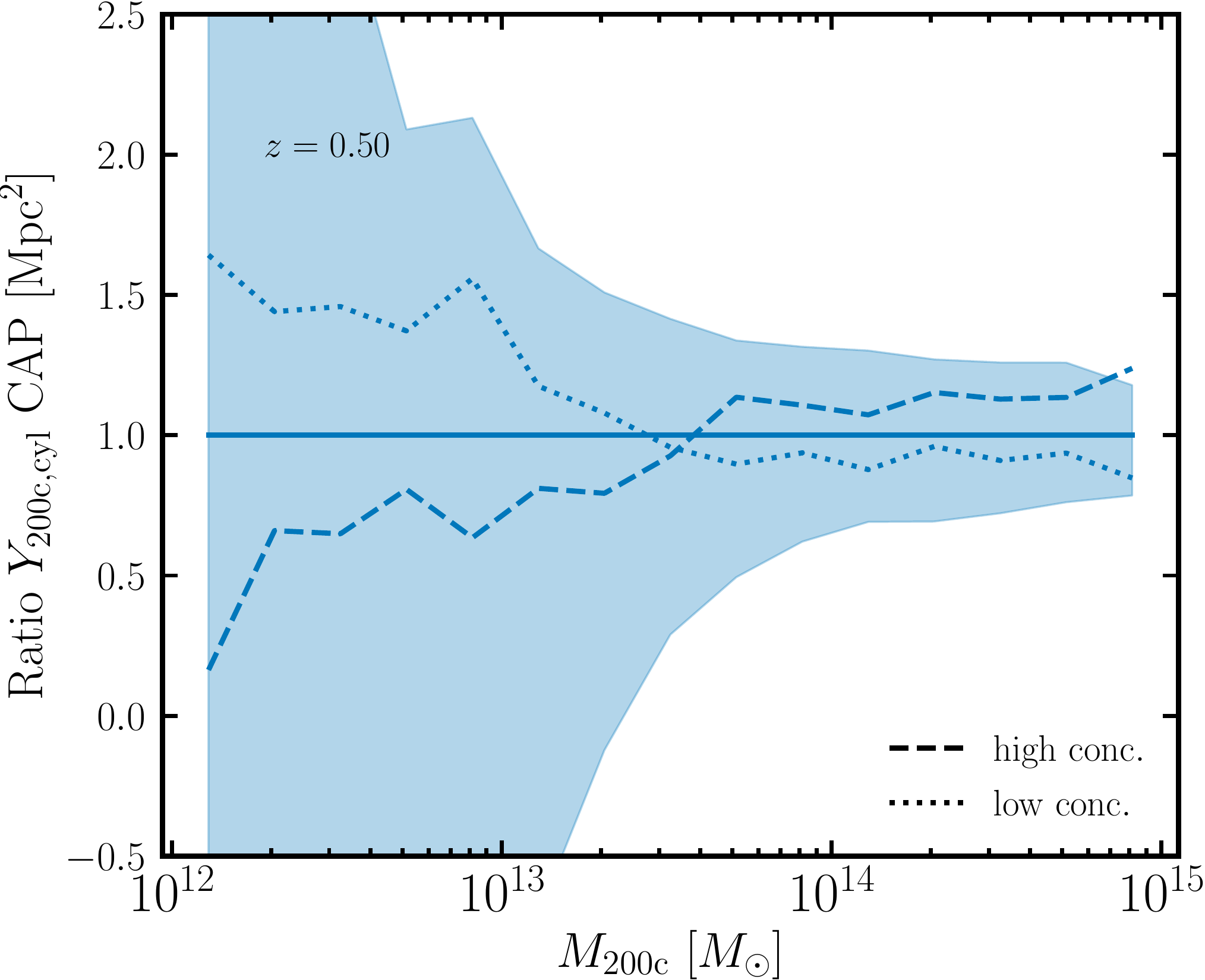}//
    \includegraphics[width=0.45\textwidth]{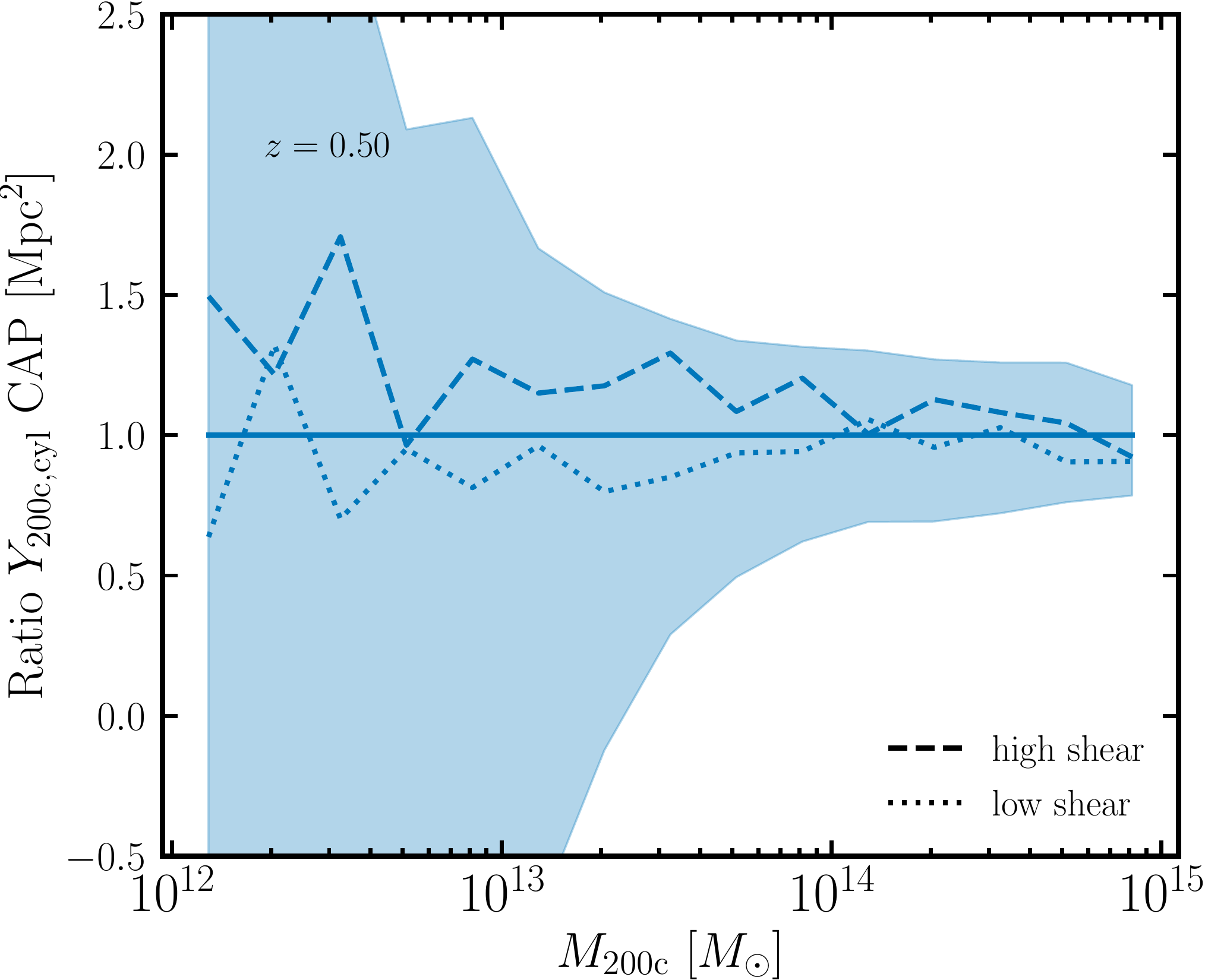}//
    \includegraphics[width=0.45\textwidth]{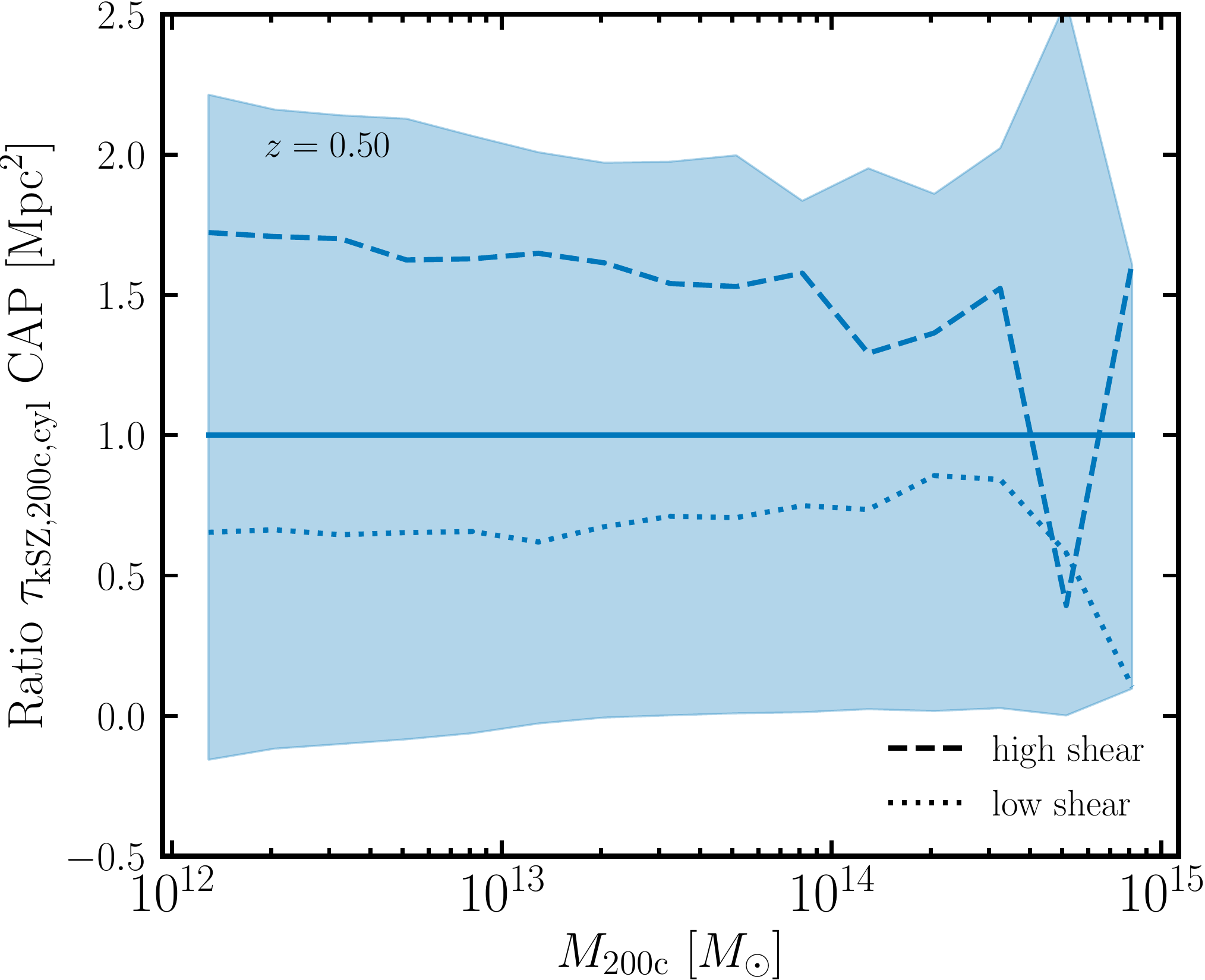}
    \caption{Dependence of the observed SZ scaling relations on concentration and shear at $z = 0.5$\teal{, shown as the ratio of the scaling relation to its mean}. The band denotes the standard deviation. The dashed and dotted curves correspond to the high and low values of either the shear or concentration. We see a mild dependence on the concentration (defined using MTNG740-DM): in the case of high-mass halos, for which gravity dominates over feedback, concentration correlates positively with the strength of the SZ signal. This trend reverses at low halo masses, for which high concentration implies more active black hole accretion and thus more violent expulsion of the baryons. The effect of shear on the scaling relations is two-fold: on one hand, larger values of shear correspond to more clustered regions and thus a stronger SZ signal from the two-halo term; on the other, anisotropic pulling from massive clusters leads to an asymmetric velocity distribution within and outside the halo, which affects the kSZ-inferred optical depth (but not as much for velocity-independent quantities such as $Y$, shown in the middle panel).}
    \label{fig:conc_env}
\end{figure}

In Fig.~\ref{fig:SBPL}, we comment on the large scatter observed in the scaling relations for both the intrinsic signal (i.e., measurements in a sphere) as well as the observed signal (i.e., measurements in a cylinder with a CAP filter). In the case of the intrinsic scatter, which is larger for lower-mass halos, we can seek an explanation in the formation history of the halo and its internal properties, which would inevitably affect its gas properties, and thus the SZ observables. On the other hand, the large scatter we see in the observed relations can be linked to the two-halo term, which is determined by the halo surroundings and is thus highly sensitive to halo properties correlated with the clustering. In this section, we examine how much of the scatter in the relations can be captured by considering the dependence of the signal on archetypal internal and external halo properties.

In addition, the sample of galaxies around which we measure the SZ effect can be influenced by selection effects related to the halo properties, an effect known as `galaxy assembly bias'. This effect refers to the discrepancy between the true distribution of galaxies and the distribution inferred from the dark matter halos using present-day halo mass \citep[e.g.,][]{2007MNRAS.374.1303C}. It has been argued that additional halo properties such as the host halo formation time, local environment, concentration and spin need to be considered to describe the clustering on small and intermediate scales correctly. 
Recent studies have shown that the population of red massive galaxies, which is often targeted in cross-correlation analysis with the CMB, is affected by galaxy assembly bias \citep[see e.g.,][for studies performed with simulations and observations]{2018MNRAS.477.4348M, 2020MNRAS.493.5506H, 2021MNRAS.502.3242X, 2021MNRAS.501.1603H, 2021MNRAS.502.3582Y}, rendering the study of assembly bias highly relevant to SZ science.

We review the definitions of two of the most relevant halo properties, concentration and shear, in an effort to capture the bulk of the effect of properties beyond halo mass on the scaling relations:
\begin{itemize}
    \item {\bf Concentration:} 
    Concentration has been linked with both halo accretion history and halo clustering \citep[e.g.,][]{1997ApJ...490..493N, 2006ApJ...652...71W, 2014MNRAS.441..378L, 2016MNRAS.460.1214L,2001MNRAS.321..559B,2014MNRAS.441..378L,2015ApJ...799..108D,2014MNRAS.441.3359D,2018MNRAS.474.5143M}. In this work, we adopt the following definition of concentration:
    \begin{equation}
    c = V_{\rm max}/V_{\rm vir},
    \end{equation}
    where $V_{\rm max}$ is the maximum circular velocity of the halo at the redshift of interest, and $V_{\rm 200c} \equiv \sqrt{GM_{\rm 200c}/R_{\rm 200c}}$, with both quantities coming from the dark-matter-only counterpart of the simulation, MTNG740-DM, \teal{excluding halos for which the matching failed (< 3\%)}. We make this choice so that our findings can be tied to the underlying dark-matter field and thus applicable for studies involving pure $N$-body simulations.
    
    \item {\bf Shear:} Our procedure for obtaining the shear is identical to the one adopted in \citet{2022arXiv221010072H}. To calculate the local ``shear'' around a halo, we first compute a dimensionless version of the tidal tensor, $T_{ij}\equiv \partial^2 \phi_R/\partial x_i \partial x_j$, where $\phi_R$ is the dimensionless potential field calculated using Poisson's equation, $\nabla^2 \phi_R = -\rho_R/\bar{\rho}$, and $R$ is the smoothing scale. We then calculate the tidal shear, $q^2_R$, as:
    \begin{equation}
    q^2_R\equiv \frac{1}{2} \big[ (\lambda_2-\lambda_1)^2+(\lambda_3-\lambda_1)^2+(\lambda_3-\lambda_2)^2\big] \, ,
    \end{equation}
    where $\lambda_i$ are the eigenvalues of $T_{ij}$. The ``shear'' measures the amount of anisotropic pulling due to gravity at a given point in space. We record the shear at the position of each halo center, adopting an adaptive smoothing scale ranging between $R = 1.1$ and $1.5 \, h^{-1} {\rm Mpc}$, such that the $R$ is always larger than the halo radius.
\end{itemize}

In Fig.~\ref{fig:conc_env}, we show the response of the tSZ and kSZ observed signals to concentration and shear\teal{, as ratios with respect to the mean scaling relation}. The blue band corresponds to the standard deviation as a function of halo mass, whereas the dashed and dotted lines correspond to the high and low values, respectively, of concentration (top panel) and environment (bottom two panels) at fixed halo mass. We note, as discussed in Section \ref{sec:sbpl}, that we opt to compute the arithmetic mean, as this is more akin to how observational analysis is performed, as it cancels additive noise terms. Although we focus on $z = 0.5$, the result is qualitatively similar across all redshifts considered in this work. \teal{We emphasize that here we show the observed SZ quantities (after applying a CAP filter), so most of the noise comes from intervening structure along the line-of-sight.}

Naively, one would expect that halos with a high value of their dark-matter concentration would also on average have a higher tSZ signal, as the baryons, which roughly follow the dark matter due to gravity, would be more concentrated within $R_{\rm 200c}$ rather than found on the outskirts. Indeed, on the high-mass end of the $Y-M$ panel, we see that this conjecture holds true. However, as we approach lower halo masses, below $\log(M_{\rm 200c}) \lesssim 13.8$, we notice that the trend reverses. It is curious that this scale corresponds roughly to the break in the power law where we see deviations from self-similarity (see Section~\ref{sec:sbpl}). We hypothesize that similarly to the break in the power law, this effect is caused by the tug of war between astrophysical feedback, which tends to push baryons out, and gravity, which tends to pull them towards the halo center. In particular, we argue that halos with high concentration tend to have more actively accreting black holes at their cores, which are more likely to go into AGN-mode and eject gas out of the halo radius, $R_{\rm 200c}$. We find evidence for this in the baryon radial profiles \teal{in that in the low-mass regime, high-concentration halos deposit more of their gas on the outskirts compared with their low-concentration counterparts}, but we defer a more detailed study to future work. In addition, we comment that this phenomenon is driven by the one-halo term, as studying the dependence of the intrinsic quantities, $Y_{\rm 200c, sph}$ and $\tau_{\rm 200c, sph}$, on concentration yields a qualitatively similar (and more evident) result. \teal{A more direct way of tying the AGN accretion rate to the assembly history of the halo might be through an alternative parametrization such as the one in \citet{2015ApJ...806...68L}}. Finally, previous works \citep[e.g.,][]{2023PNAS..12002074W,2022MNRAS.517..420L} find that across all mass scales, high concentration halos have a higher $Y$ signal, unlike the mass-dependent trend we observe. \teal{We attribute this apparent discrepancy to the definition of halo concentration and the difference in halo finder, and leave the study of alternative assembly history definitions for follow-up work.}

In the bottom panel, we explore the response of the kSZ-inferred $\tau-M$ relation to the environment of the halo, through the parameter `shear.' \teal{We opt not to display the dependence on concentration, as the result is qualitatively very similar to that of $Y$.} We see a strong response to shear across all halo masses, suggesting that halos experiencing a high-level of anisotropic pulling will have on average higher kSZ-inferred optical depth, $\tau_{\rm kSZ, 200c, cyl}$, than their low-shear counterparts. To understand this result, we mention several relevant findings, some of which are not shown in the figure: 1) This finding holds true even if we confine our kSZ-inferred optical depth to the sphere rather than the cylinder, $\tau_{\rm kSZ, 200c, sph}$. 2) On the other hand, were we to repeat this exercise with the intrinsic optical depth and $Y$ signal, $\tau_{\rm 200c, sph}$ and $Y_{\rm 200c, sph}$, we would find negligible dependence on shear. 3) A final piece of the puzzle comes from considering the quantities $\tau_{\rm 200c, cyl}$ and $Y_{\rm 200c, cyl}$, both of which we find to have a qualitatively similar, but much weaker dependence on shear to $\tau_{\rm kSZ, 200c, cyl}$ (the latter of which is shown in the middle panel). 

We hypothesize that halo environment, characterized by `shear,' has little effect on the one-halo term of the electron pressure and number density, but has a substantial effect on the two-halo term, as it correlates strongly with the presence of massive nearby structures that contribute to the line-of-sight signal. In the case of the kSZ-inferred $\tau$, the signal depends not only on the electron number density, but also on the velocity distribution within (and outside) the halo. We argue that in the presence of a massive nearby cluster, the halo experiences a large amount of gravitational pulling, which leads to an asymmetric spatial and velocity distribution of the electrons: the parts of the halo closer to the massive cluster have on average higher velocities than its center-of-mass (though pointing in a coherent direction) as well as higher densities (leading to an anisotropic gas distribution), \teal{resulting in a large gradient in the bulk velocity}. \teal{We note that to first order in linear theory, there should be no correlation between velocity and shear, but it appears that in the small-scale non-linear regime, this assumption breaks down.} Both of these effects conspire to yield an increase in the inferred optical depth from kSZ measurements.

\begin{figure}
    \includegraphics[width=0.47\textwidth]{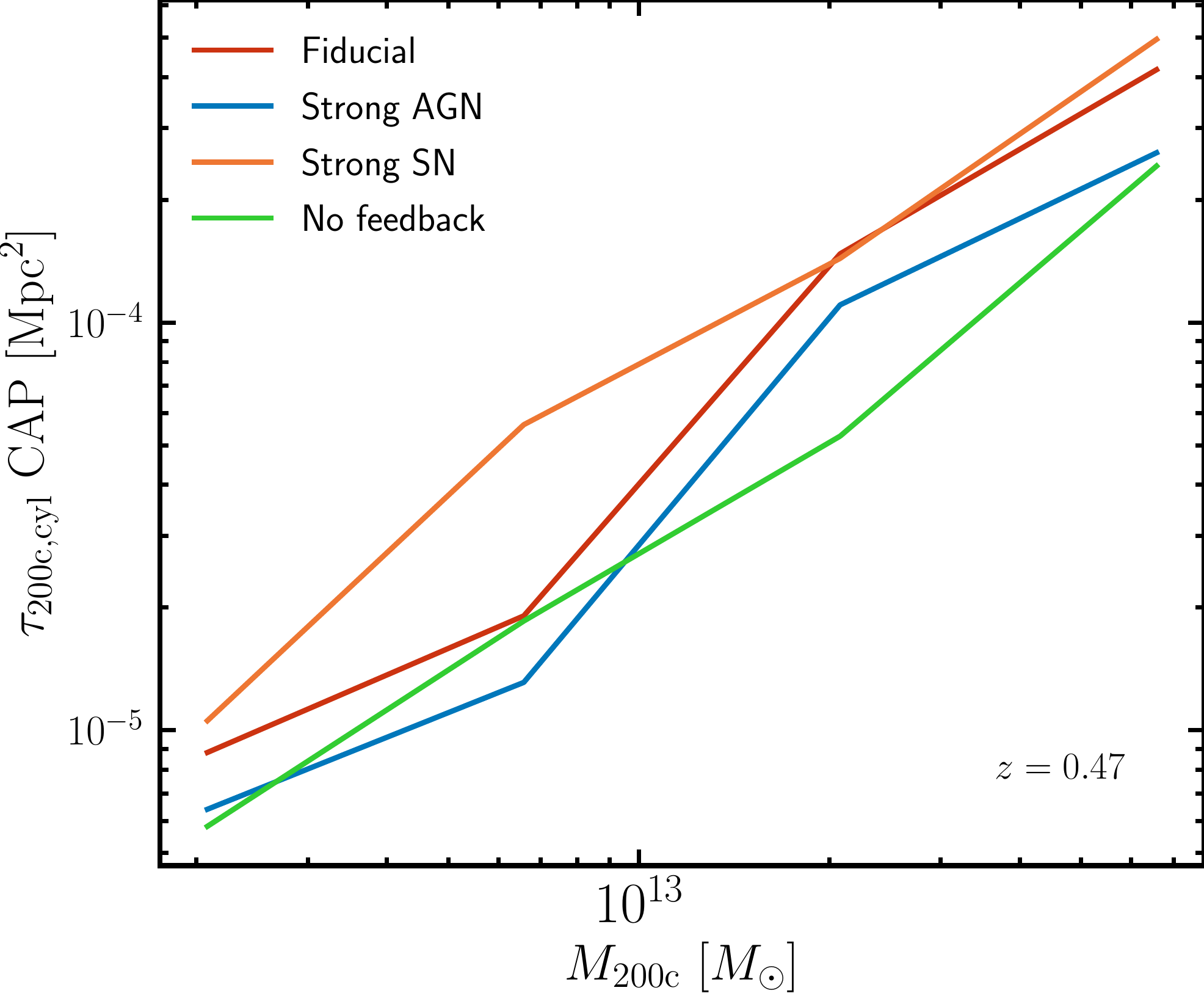}\\
    \includegraphics[width=0.47\textwidth]{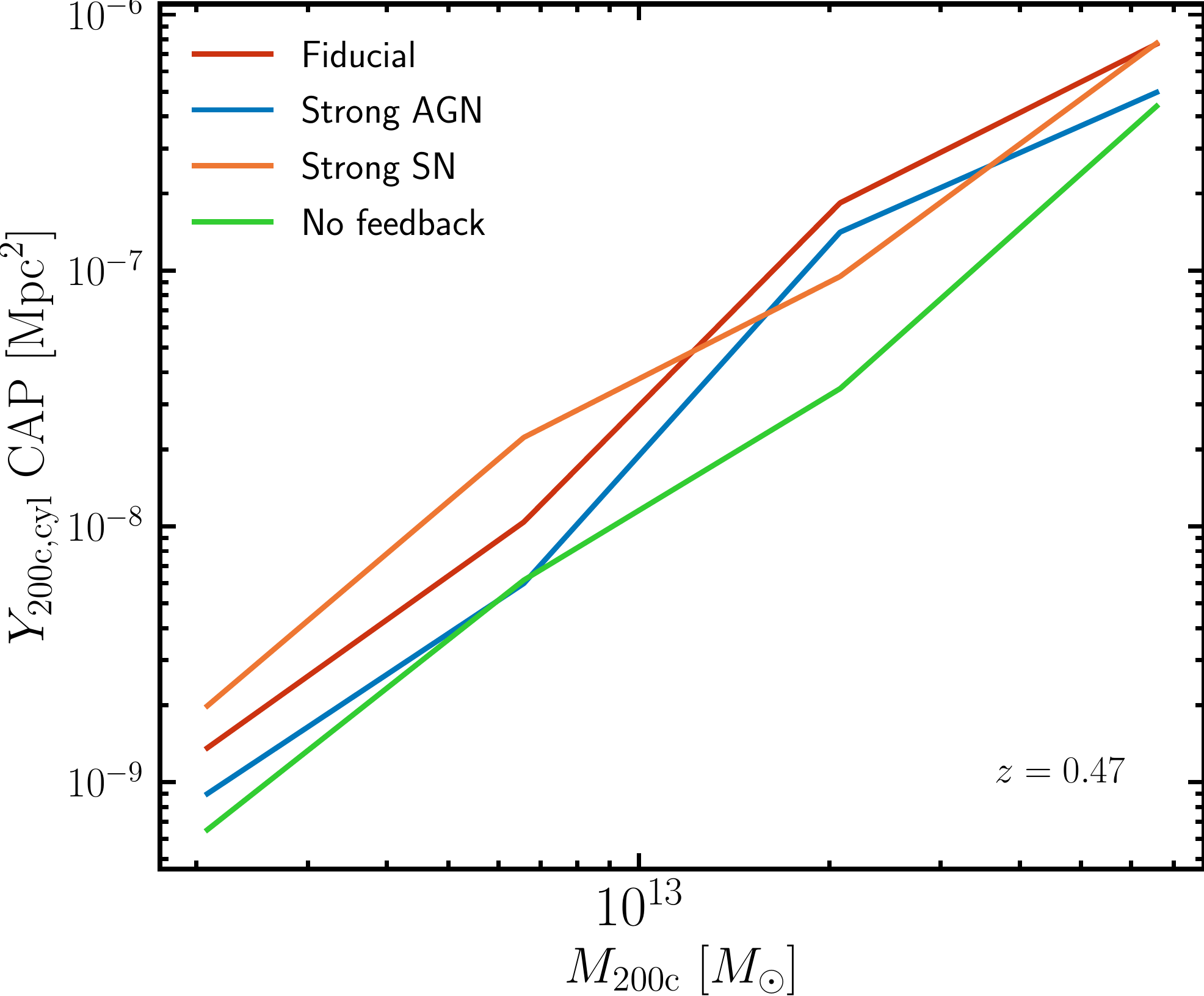}
    \caption{Scaling relations of the optical depth (top panel) and the integrated Compton-y signal (bottom) for the four CAMELS `extreme feedback' simulations at $z \approx 0.5$. The quantities are measured in a cylinder and a CAP filter is applied. The extreme SN box shows largest differences from the fiducial at low masses, while the extreme AGN deviates the most at high masses, as expected. The no-feedback box follows an almost perfect power law, as the gas dynamics is dominated by gravity. We quantify the error due to the subgrid modeling at 30\% for $M_{\rm 200c} \sim 10^{13} \, {\rm M}_\odot$ and 15\% for the highest-mass halos present in the CAMELS suite.}
    \label{fig:camels}
\end{figure}

\subsection{Dependence on the astrophysical feedback model}
\label{sec:camels_sub}

As discussed in Section~\ref{sec:sbpl}, the fits to the scaling relations are strongly dependent on the assumed small-scale astrophysics. In particular, relevant to SZ are the effects of supernova and AGN feedback, as they are the most effective processes for expelling baryons from groups and clusters. Unfortunately, large hydro simulations are extremely expensive to run with different subgrid models, so to quantify these effects, we resort to using the much smaller-volume CAMELS simulations (see Section \ref{sec:camels}). We make use of the four `extreme feedback' boxes corresponding to: 1) the fiducial subgrid model, 2) a model with strong SN feedback, 3) a model with strong AGN feedback, and 4) a model with no feedback. In some ways, this is a conservative choice, as these boxes do not recover many of the observable quantities that MTNG and IllustrisTNG are tuned to match, so the feedback levels in these simulations are likely not realistic. On the other hand, the hydrodynamical code used to generate the `extreme' boxes is the same as that of MTNG and IllustrisTNG, and a more robust test would involve exploring how the SZ observables vary for different hydrodynamical codes \citep[see][]{2023PNAS..12002074W}. Planned for the near future\footnote{As part of the Learning the Universe (LtU) initiative.} is the creation of small zoom-in simulations implementing a wide range of feedback models into hydrodynamical and semi-analytic models, and we defer a more detailed study to when such resources become available. Finally, we note that since the CAMELS boxes are much smaller, they are lacking in massive halos above several times $10^{13} \, {\rm M}_\odot$, so our CAMELS study is limited to small- and intermediate-mass halos.

Shown in Fig.~\ref{fig:camels} are the observed $Y-M$ and $\tau-M$ scaling relations for the four `extreme' simulations at $z = 0.5$. The model with no feedback yields an almost constant-slope relation for both pairs of observables, confirming the claim that any deviation from self-similarity is likely the result of feedback especially at $M_{\rm 200c} \approx 10^{13} \, {\rm M}_\odot$. On the other hand, the curve corresponding to the simulation with large SN feedback differs the most from the fiducial for small-mass halos, suggesting that at these mass scales, supernova feedback has a large effect on SZ observables. This statement is reversed as we go to higher masses. We also note that these statistics come from $\sim$50 objects, so there is additional scatter from the low number of samples. A conservative estimate of the effect of feedback on the inferred scaling relations is that the $Y-M$ and $\tau-M$ relations vary by about 30\% at $10^{13} \, {\rm M}_\odot$, with that number decreasing to about 15\% as we go to highest masses we consider.

\subsection{Measurements around stellar-mass-selected galaxies}
\label{sec:DESILRG}

In this Section, we make a more direct connection with observations by measuring the tSZ and kSZ effect around galaxies as opposed to halos, which are not directly observable. When making measurements of the SZ effect, a preferred option is to stack the CMB signal on a roughly homogeneous population of galaxies. Since the SZ signal is proportional to the number of electrons contained in the cluster, a sensible choice for a galaxy sample is the population of massive red galaxies such as the SDSS CMASS (`constant mass') sample \citep{2016MNRAS.460.1173R} and the luminous red galaxies (LRGs) of Dark Energy Spectroscopic Instrument \citep[DESI][]{2016arXiv161100036D}. 

Similarly to \citet{2022arXiv221010059H}, here we select galaxies by applying a stellar mass cut to the subhalos in MTNG such that their number density is $n_{\rm gal} = 1 \times 10^{-3} \ [\hMpc]^{-3}$, roughly matching the expected number density of red galaxies in modern surveys at that redshift. The corresponding minimum stellar mass is $7.4 \times 10^{10} \ \hMsun$, and the satellite fraction is 21\%.

Since the shapes of the tSZ and kSZ profiles are unknown, a matched filter cannot be reliably employed to recover the shape of the baryon profiles. Therefore, alternatives such as the CAP filter with a varying aperture radius $\theta_d$, which is agnostic about the profile shape, are often adopted. As detailed in \cite{Schaan:2020}, applying that filter to a temperature map $\delta T$ yields: 
\begin{equation}
\label{eq:ap}
\mathcal{T}(\theta_{d})=
\int {\rm d}^2\theta \, \delta T(\theta) \, W_{\theta_{d}}(\theta) \,.
\end{equation}
where the filter $W_{\theta_{d}}$ is chosen as:
\begin{equation}
W_{\theta_{d}}(\theta) =
\left\{
\begin{aligned}
1& &  &\text{for} \, \theta < \theta_d \,, \\
-1& &  &\text{for} \, \theta_d \leq \theta \leq \sqrt{2}\theta_d \,, \\
0& & &\text{otherwise}. \\
\end{aligned}
\right.
\end{equation}
\teal{Note that many of the \textit{Planck} team analyses use a similar compensated filter \citep[e.g.,][]{2013A&A...557A..52P,2013A&A...550A.129P,2014A&A...571A..20P}.} This corresponds to adding the integrated temperature fluctuation in a disk with radius $\theta_d$ and then subtracting the integrated temperature fluctuation in a concentric ring of the same area as the disk, at each radial bin. Similarly to \cite{Schaan:2020}, we vary the size of the radial bins, $\theta_d$, between 1 and 6 arcmin. At $z = 0.5$, where we study the LRG signal, this corresponds to roughly 0.5 to 3 times the host halo radius for the galaxies in our median stellar mass bin. A convenient property of the CAP-filtered profiles is that they behave similarly to a cumulative density/pressure profile for large radii. 

\begin{figure}
    \includegraphics[width=0.48\textwidth]{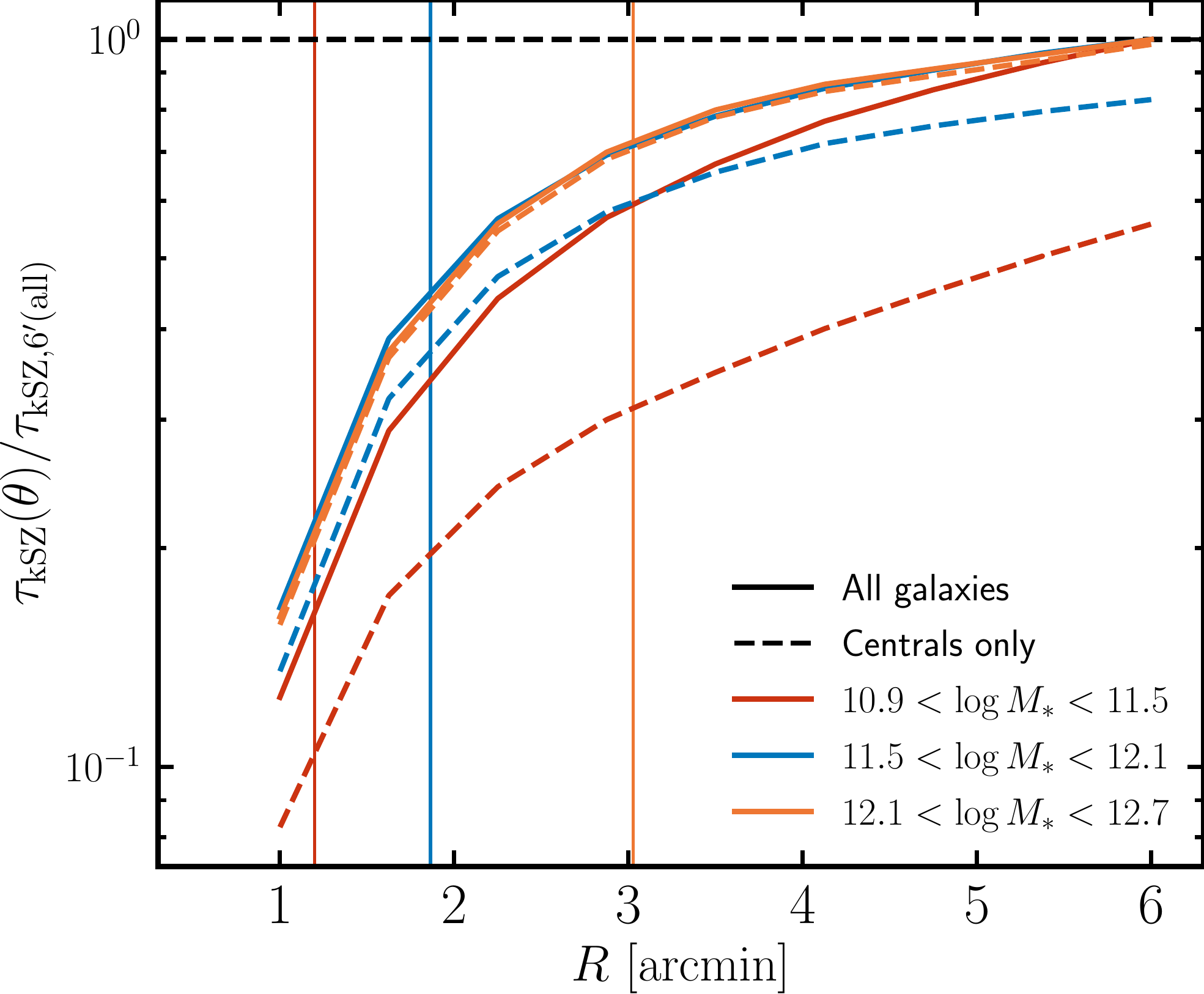}
    \caption{Compensated Aperture Photometry (CAP) measurements of the kSZ signal at $z \approx 0.5$ for a stellar-mass selected sample of galaxies. The galaxies are split into three stellar mass bins (mimicking cuts in luminosity). We study separately the all-galaxies (solid) and the centrals-only (dashed) populations so as to quantify the effect of `miscentering.' Miscentering affects the lowest-mass population the most, changing the shape of the profile and amplitude relative to the centrals-only curve. The effect of miscentering on the middle bin, which is closest to the median stellar mass of galaxy surveys such as CMASS and DESI, is about 20\%, warranting careful modeling of the effect when analyzing observations.}
    \label{fig:CAP}
\end{figure}

In Fig. \ref{fig:CAP}, we show the stacked kSZ-inferred optical depth profiles as a function of distance from the galaxy center. We bin the galaxies in stellar mass, as stellar mass is a good proxy for galaxy luminosity, which is typically the observable galaxy property used to bin SZ measurements \citep{2021PhRvD.104d3503V}. In order to compare the shape of the profiles self-consistently, we normalize the profiles by the total optical depth within 6 arcmin of the `all-galaxies' sample for each stellar mass bin.
If we opted not to normalize, we would find that the profiles follow the expected trend: massive galaxies, which tend to live in massive halos, have large optical depth. Overall, we note that the galaxies in the lowest stellar-mass bin have the steepest profiles, whereas the galaxies in the two most massive bins have almost indistinguishable profiles. We attribute this to the effects of miscentering, which is the result of not being able to discriminate centrals from satellites in the data, and the two-halo term, which has a relatively larger contribution for less massive clusters as previously discussed (see Fig.~\ref{fig:tau_M200c}).

Since in simulations we can distinguish between satellites and centrals, we can study the effect of miscentering on the SZ profiles by splitting the sample in each stellar mass bin into `all galaxies' and `centrals-only.' The role miscentering appears to play on the signal is two-fold. On one hand, the central region of the halo, which dominates the signal, is now offset from the satellite center used in the stacking, which yields an anisotropic signal and leads to a steepening of the radially averaged profile that is typically studied. On the other hand, the amplitude of the SZ effect is lower, as at fixed stellar mass, centrals tend to live in less massive halos (and thus have a lower SZ signal) compared with satellites. There are barely any satellites with stellar masses above $10^{12} \, {\rm M}_\odot$, so the curve showing the centrals-only profile looks almost identical to the all-galaxies one. We note that for galaxies of stellar mass $\sim$$10^{12} \, {\rm M}_\odot$, which is representative of the median mass of surveys such as CMASS and DESI, miscentering appears to have a 20\% effect on the inferred profiles, as the satellite fraction is relatively low compared with the lowest mass bin. Given the expected precision of these measurements, it is highly advisable to model this effect in observational analyses, for example through Halo Occupation Distribution (HOD) models (Eg. \cite{Zheng:2004id}).

\section{Summary and conclusions}
\label{sec:conc}

Cosmology has entered an era in which high-resolution large-volume hydrodynamical simulations can be used to inform observations of the large-scale structure and their cross-correlations with early-Universe probes such as the Cosmic Microwave Background (CMB). Conversely, the joint analysis of early- and late-time observables yields subpercent constraints on measurable quantities that have hitherto been unconstrained, providing an independent test of hydrodynamical codes and their subgrid physics implementations. In this work, we study the thermal and kinematic Sunyaev-Zel'dovich (SZ) effects in the largest-to-date hydrodynamical simulation, MillenniumTNG. In particular, we focus on interpreting the integrated Compton-y parameter, $Y$, and optical depth, $\tau$,  offering useful insight for future observational analyses. We next highlight our key findings.

We first explore the various contributions to the $Y$ and $\tau$ signal in Fig.~\ref{fig:Y_M200c} and Fig.~\ref{fig:tau_M200c}. We find that there is a substantial contribution to the observed signal from random uncorrelated structure along the line-of-sight (especially for small-mass halos), which can be efficiently removed by utilizing a Compensated Aperture Photometry (CAP) filter. We contrast that with a cleaning procedure, in which we subtract the SZ signal from random points in the sky, finding that this method overshoots by removing too much of the signal. The remaining signal is made up of the combination of the one- and two-halo terms, stemming from the halo of interest and clustering from nearby halos, respectively.

Apart from the random contribution, Fig.~\ref{fig:tau_M200c} also compares the various $\tau$-related quantities. We find that the integrated cylindrical~$\tau$, which is the relevant quantity in FRB analysis, can be as much as an order of magnitude larger than the kSZ-inferred $\tau$ and the spherical~$\tau$, the latter of which is close to the quantity measured in X-ray experiments. Intriguingly, we find that the kSZ-inferred $\tau$, on average, is most akin to the CAP-filtered cylindical~$\tau$ and approaches the intrinsic $\tau$ for large halo masses. We explain this by noting that due to the cancellation of velocities along the line-of-sight, the kSZ effect is most sensitive to structures in a short cylinder of length several 10s of Mpc around the halo of interest. The size of the cylinder is determined by the decorrelation length of the radial velocity field, which decorrelates faster than matter.

We next provide fits to the intrinsic and observed scaling relations in Fig.~\ref{fig:SBPL} and Table~\ref{tab:SBPL} adopting the `smoothly broken power law' formalism. We observe that the slope deviates from self-similarity for halos below $M_{\rm 200c} \lesssim 10^{14} \, {\rm M}_\odot$, echoing previous findings in both simulations and observations regarding the effect of AGN and supernova feedback. The break is obscured to a degree when considering the cylindrical CAP-filtered quantities, for which the SZ contribution external to the halo adds noise to the relation and changes the slope at lower halo mass. We note that the slope is nearly constant for $Y-\tau$ as both quantities are affected by the same astrophysical processes. It is noteworthy that the scatter in the observed $Y-\tau$ relation is substantial and results from the fact that the kSZ-inferred $\tau$ receives large noise contributions from the line-of-sight velocities, which can fluctuate by quite a bit around the bulk radial velocity of the halo. 

We explore the source of the scatter in both the observed and the intrinsic scaling relations in Fig.~\ref{fig:conc_env}. While we find that the intrinsic $Y$ and $\tau$ of a cluster are largely insensitive to its environment, they do depend on the halo concentration, which we define using the dark-matter-only counterpart of MTNG740, so as to make a direct connection with the underlying dark matter field. We find that, as naively expected, in the high-mass regime, halos with higher dark-matter concentration have a stronger SZ signal, as gravity wins over baryonic feedback. However, in the low-mass regime, we see an inversion to that trend. We tie this to the increase in AGN activity in rapidly accreting systems (whose density tends to be higher in the central regions), which tends to push baryonic material more readily out of the virial radius of the halo.

The dependence of the SZ signal on environment, shown in Fig.~\ref{fig:conc_env}, through the parameter we call `shear' is quite noticeable for observed (i.e., containing the two-halo term) and kSZ-inferred quantities (i.e., depending on the radial velocity). On one hand, large values of the shear parameter are found in highly clustered regions acted upon by strong gravitational forces. As such, the contribution to the measured SZ signal from the two-halo term is much larger for these objects than their low-shear counterparts. On the other hand, the strong anisotropic pulling associated with high-shear regions also affects the velocity field of the halo and its surroundings. For example, the presence of a massive nearby cluster can induce tidal effects in smaller halos, violating their radial symmetry in phase-space and thus, affecting the intrinsic and extrinsic kSZ signal.

Simulation-inferred scaling relations can have a strong dependence on the subgrid physics model baked into the hydrodynamical code. We explore this effect on the cylindrical $Y$ and $\tau$ in Fig.~\ref{fig:camels} via the CAMELS `extreme feedback' suite, which covers a modest halo mass range between $10^{12}$ and $5 \times 10^{13} \, {\rm M}_\odot$. We find a 30\% scatter between the four boxes at $M_{\rm 200c} = 10^{13} \, {\rm M}_\odot$ and note that this is a somewhat conservative estimate, as the feedback models of boxes \texttt{EX\_1-3} are quite extreme and unlikely to correctly reflect the feedback in the real Universe. We defer a full exploration with more diverse and realistic feedback models as well as better representation of the high-mass regime to future work.

Finally, as a case study, we show the kSZ profiles of stellar-mass selected galaxies at $z = 0.5$, where samples from galaxy surveys such as SDSS and DESI peak, and split them into stellar mass (or almost equivalently, luminosity) bins. These galaxies tend to be luminous and red, and are often referred to in the literature as luminous red galaxies (LRGs). To mimic observations, we stack indiscriminately around centrals and satellites and quantify the effect of `miscentering.' We find that the most massive galaxies are hardly affected, as they are predominantly centrals. For the lowest stellar-mass bin, both the amplitude and the shape are affected, as the satellites, whose fraction is now higher, live in relatively more massive halos than centrals of the same luminosity, and pick up the peak of the SZ signal at a small offset determined by their distance to the center. Most representative of current and near-future surveys is the second mass bin, which seems to retain the correct shape, but suffers a 20\% mitigation when considering the centrals-only population. Given the projected unprecentedly small error bars expected for these measurements from upcoming experiments, it is strongly advisable, especially for low-luminosity objects, to carefully account for the effect of miscentering in the theoretical modeling of SZ observables.

As the influx of new data flows in, theorists need to likewise step up and hone their theoretical tools in order to perform careful and robust analysis of the observed data. This paper offers a first step to interpreting and connecting SZ observations with the state-of-the-art hydrodynamical simulation MTNG740. In the near future, we intend to make use of the full-physics light cones provided as part of the MTNG package and delve deeper into understanding the galaxy selection effects, the radial dependence of the profiles, the filter choices, the instrument noise, the role of the CMB as a contaminant, and the connection between baryons and dark matter, among other effects. The analysis of CMB secondary anisotropies promises exciting new developments in the realm of astrophysics and cosmology in the next few years.

\section*{Acknowledgements}

We thank Nick Battaglia, Eve Vavagiakis, Shivam Pandey, Emmanuel Schaan, Shy Genel, Daisuke Nagai, and members of the Simons Collaboration on ``Learning the Universe'' for many enlightening discussions. 
SF is supported by the Physics Division of Lawrence Berkeley National Laboratory. SB is supported by the UK Research and Innovation (UKRI) Future Leaders Fellowship [MR/V023381/1].

\section*{Data Availability}

 The data underlying this article will be shared upon reasonable request to the corresponding 
authors. All MTNG simulations will be made publicly available in 2024 at the following address: \url{https://www.mtng-project.org}.



\bibliographystyle{mnras}
\bibliography{example} 




\appendix

\bsp	
\label{lastpage}
\end{document}